\documentclass[sigconf]{acmart}

\usepackage{multirow,graphicx}
\usepackage{wrapfig}
\usepackage{array} 
\usepackage{amsfonts}
\usepackage{algpseudocode}
\newcolumntype{H}{>{\setbox0=\hbox\bgroup}c<{\egroup}@{}} 
\usepackage{url}

\usepackage{hyphenat}

\newtheorem{mydef}{Definition}

\usepackage{subcaption}
\usepackage{balance}

\usepackage[lined,ruled,linesnumbered,noend]{algorithm2e}
\usepackage{amsmath}
\usepackage{booktabs,pifont}
\usepackage[utf8]{inputenc}
\usepackage{tcolorbox}
\usepackage{enumitem}
\tcbuselibrary{skins, breakable}

\newtcolorbox{boxA}{
  breakable,
  colback=gray!10,
  colframe=black!60,
  arc=2mm,
  left=6pt,
  right=6pt,
  top=6pt,
  bottom=6pt
}

\usepackage{multirow}
\usepackage{xcolor}

\usepackage{threeparttable}

\AtBeginDocument{%
  }
    
\acmYear{2026}\copyrightyear{2026}
\setcopyright{cc}
\setcctype[4.0]{by}
\acmConference[ASIA CCS '26]{ACM Asia Conference on Computer and Communications Security}{June 1--5, 2026}{Bangalore, India}
\acmBooktitle{ACM Asia Conference on Computer and Communications Security (ASIA CCS '26), June 1--5, 2026, Bangalore, India}
\acmDOI{10.1145/3779208.3785388}
\acmISBN{979-8-4007-2356-8/26/06}

\begin{document}

\title{Original Sin of {\tt npm}: A Study on Vulnerability Propagation in JavaScript Dependency Networks}

\author{Michael Robinson}
\affiliation{%
 \institution{Data61, CSIRO}
 \country{Australia}}
 \email{michael.robinson@data61.csiro.au}
 
\author{Sajal Halder}
\orcid{0000-0002-0965-6255}
\affiliation{%
  \institution{Data61, CSIRO \& RMIT University}
\country{Australia}}
\email{sajal.halder@rmit.edu.au}

\author{Muhammad Ejaz Ahmed}
\affiliation{%
 \institution{Data61, CSIRO}
 \country{Australia}}
 \email{ejaz.ahmed@data61.csiro.au}

\author{Muhammad Ikram}
\affiliation{%
 \institution{Macquarie University}
 \country{Australia}}
 \email{muhammad.ikram@mq.edu.au}

 \author{Seyit Camtepe}
\affiliation{%
 \institution{Data61, CSIRO}
 \country{Australia}}
 \email{seyit.camtepe@data61.csiro.au}

 \author{Hyoungshick Kim}
\affiliation{%
 \institution{Sungkyunkwan University}
 \country{South Korea}}
 \email{hyoung@skku.edu}

\renewcommand{\shortauthors}{Michael Robinson, et al.}

\begin{abstract}
In software development, the widespread adoption of open-source software has led to a proliferation of reusable modules and package libraries such as the Node Package Manager ({\tt npm}). While this approach can significantly reduce development time, it also introduces the risk of vulnerability inheritance where a dependent \textit{parent package} inherits vulnerabilities from a referenced \textit{dependency package}. Understanding vulnerability propagation is essential for assessing how vulnerabilities spread across components of a software package. This supports more accurate impact analysis and enhances threat detection and mitigation. There is a need for frameworks that capture propagation across multiple layers and dynamic environments, along with comprehensive simulation tools that reflect real-world software systems. In this paper, we investigate how a small number of vulnerable JavaScript packages contribute to the creation of a disproportionately large number of vulnerable packages.  This paper presents insights from 1,515 reported vulnerabilities gathered from a custom-built vulnerability database containing 1,077,946 JavaScript packages sourced from `{\tt npm-follower}' and their associated dependency networks. Dependency networks were constructed using the {\tt deps.dev} API, with vulnerabilities identified by parsing package names and version numbers through the Google Open Source Vulnerability API.

Our findings reveal that 61.30\% (660,748) of packages are reliant on one or more dependency packages, and 21.60\% (232,836) of total packages have at least one known vulnerability throughout their dependency networks--of which most (42\%) are of High severity. Moreover, we found that the number of vulnerabilities and affected packages on {\tt npm} since 2015 has increased linearly--with $R^2$ values of 0.97 and 0.96, respectively. We also found that it takes, on average, approximately 4 years and 11 months to fix a vulnerable package from when the first vulnerable version is published on {\tt npm} -- although publication times of vulnerabilities occur approximately 19 days after a fix is available.  Finally, we observe a high concentration of frequently present vulnerabilities throughout dependency networks, with the top-7 most frequent vulnerabilities accounting for 25\% of vulnerability cases and the top-23 most frequent accounting for 50\%.  Based on these findings, we propose recommendations for developers and package managers to mitigate the threat and occurrence of vulnerabilities within the {\tt npm} dependency network and the broader software repository community.
\end{abstract}

\ccsdesc[500]{Security and privacy~Software and application security}
\keywords{npm, JavaScript, software dependencies, vulnerability propagation}

\maketitle

\vspace{-1.5mm}\section{Introduction}
\label{sec:intro}
Software repositories serve as essential components for managing both private and public software projects. Private repositories are accessible and modifiable only by authorized users, typically owned by individuals or companies. Public repositories, on the other hand, are open to all users. Within publicly accessible packages is open-source software (OSS), characterized by decentralized development that allows numerous developers to contribute to and modify the software. This collaborative approach offers significant benefits, such as accelerated development times and reduced costs for firms that would otherwise need to develop and maintain software in-house \citep{hoffman2024value}. Accordingly, open-source development has gained significant traction in recent years, with one study revealing that, as of April 2024, 96\% of 1,067 codebases gathered across various package managers contained open-source code \citep{synopsys2024}.

Given the prevalence of OSS, it has become standard practice to incorporate functions from pre-existing OSS projects available within a repository ecosystem, further accelerating development times and minimizing costs~\citep{Ayala2013,giordano2024adoption,hoffman2024value}. Consequently, modern OSS development has evolved into a complex web of interlinked dependencies between many packages, significantly increasing the risk of inheriting vulnerabilities through dependency networks arising due to potentially malicious actors or programming errors. One of the largest repositories for such packages is the {\tt npm}, which is essential for developing JavaScript (JS) based applications and hosts nearly 4 million JS packages\footnote{The `{\tt npm}-follower' dataset contained 3,946,361 unique packages at the time of download (as of 5 September 2023). This number extends to 36,432,989 when considering each unique version of each package.}.  The {\tt npm} is consequently integral to the JS ecosystem, powering numerous applications and services that are foundational to modern web development through millions of packages that are reused in countless web applications, making it a prime target for malicious activity and programming errors that introduce vulnerabilities \citep{zerouali2018empirical}.

Identifying and mitigating vulnerabilities in OSS is, therefore, crucial to maintaining software security.  A study by Liu et al. highlights how critical vulnerabilities, if left undetected or unpatched, can propagate across software ecosystems, emphasizing the importance of prompt vulnerability discovery \citep{liu2022demystifying}.  Real-world incidents, such as the Heartbleed vulnerability in OpenSSL or the more recent Log4Shell vulnerability in Apache Log4j, demonstrate the potential impact of unpatched vulnerabilities \citep{carvalho2014heartbleed, hiesgen2022race}.  Recent advancements in vulnerability detection tools, such as `V1SCAN'~\citep{woo2023v1scan}, leverage code classification techniques to identify vulnerabilities in reused open-source components before they become widespread.  However, a major challenge remains: understanding how vulnerabilities propagate through {\tt npm} dependency networks and identifying which packages are most at risk due to outdated dependencies.

Understanding software dependency networks is crucial for assessing security risks in modern package ecosystems. The structure and evolution of the {\tt npm} dependency network plays a critical role in our study, as its rapid expansion and interconnectivity make it particularly susceptible to security vulnerabilities.  Previous research has shown that the {\tt npm} ecosystem is characterized by \textit{superlinear} growth in the number of packages and updates, with approximately 81.3\% of packages depending on at least one other package by the end of August 2015 \citep{wittern2016look}. The increasingly interconnected nature of these packages through dependency networks underscores the urgency of understanding how security vulnerabilities may propagate within these networks and the need to develop methods to mitigate the spread of such vulnerabilities.

Prior work has highlighted the increasing prevalence of vulnerabilities within software repositories. For instance, Zerouali et al. revealed a sharp rise in disclosed vulnerabilities for the {\tt npm} ecosystem compared to other package management systems like RubyGems \citep{zerouali2022impact}. These findings align with the observed increase in transitive dependencies within {\tt npm}, suggesting that vulnerability risks are compounding over time \citep{kikas2017structure}.
Our work builds on this foundation by specifically examining how vulnerabilities propagate through {\tt npm}'s dependency networks. We aim to provide deeper insights into the \textit{time-to-disclosure} of vulnerabilities and the extent of \textit{outdated package dependencies}, both of which are crucial for assessing security risks within the ecosystem. To achieve this, we investigate the following research questions:

\begin{itemize}[topsep=0pt,left=0pt]
    \item \textbf{$\mathbf{RQ_1}$: What is the nature of vulnerability discovery and publication in the npm ecosystem?}
    Understanding how vulnerabilities are discovered and published is essential for assessing the responsiveness of the ecosystem to security threats. By addressing this question, we aim to determine whether vulnerabilities are increasing at a manageable rate and whether developers disclose vulnerabilities in a timely manner relative to the first known exposed version. This insight is critical for maintaining transparency and enabling the community to respond effectively to emerging security risks.
    
    \item \textbf{$\mathbf{RQ_2}$: How effective is the response to discovered vulnerabilities in the npm ecosystem?}
    The efficiency of vulnerability resolution directly impacts software security. This question evaluates whether developers are fixing vulnerabilities within a reasonable time frame after their initial publication. By identifying potential delays in the remediation process, we aim to highlight whether the \textit{publication-to-fix duration} contributes to prolonged security risks, thereby helping to improve vulnerability management strategies.
    
    \item \textbf{$\mathbf{RQ_3}$: What is the characterization of the origins of vulnerabilities and their pervasiveness within the npm ecosystem's dependency network?} 
    Dependencies play a critical role in the spread of vulnerabilities. This question examines whether vulnerabilities are disproportionately concentrated in certain packages and whether a relationship exists between the \textit{number of vulnerabilities and the number of dependencies} within a package's dependency network.
\end{itemize}

By addressing these research questions, our study contributes to a deeper understanding of the systemic security risks within the {\tt npm} ecosystem. The findings can help software maintainers and developers prioritize security measures, optimize vulnerability response times, and refine best practices for dependency management.

The remainder of the paper is organized as follows: Section \ref{priliminaries} discusses the preliminaries and formal problem definition. Section \ref{sec:methodology} describes the methodology behind our research, including the rationale for choosing the OSS environment and the creation and utility of the dependency and vulnerability databases. Section \ref{sec:study} presents our findings concerning the research questions. Section \ref{sec:discussion} synthesizes these results into actionable recommendations and further discusses the state of {\tt npm} as implied by our findings. Section \ref{sec:threats} explores adversarial threats that could undermine our findings and proposes mitigation strategies. Section~\ref{sec:rwork} discusses related work, while we conclude the paper in Section \ref{sec:conclusion}.

\section{Preliminaries \& Problem Statement }
\label{priliminaries}
\subsection{Preliminaries} 

Understanding vulnerability propagation in software ecosystems such as  the npm registry requires clear and consistent terminology. 
We therefore ground our definitions in established software  supply–chain security literature \citep{zimmermann2019small, decan2019empirical, kikas2017structure},  and in formal frameworks used by the MITRE CVE Program, the National Vulnerability Database (NVD), and the Common Vulnerability Scoring System (CVSS) \citep{cvesite2024, firstsite2024}. This ensures conceptual alignment with prior large-scale ecosystem studies  \citep{zerouali2018empirical, liu2022demystifying} and provides a rigorous basis for analysing dependency networks and vulnerability inheritance.

A \textit{vulnerability} is defined as a known security threat affecting specific releases of software packages. In software development, these vulnerabilities are commonly cataloged as Common Vulnerabilities and Exposures (CVEs)--globally unique identifiers assigned by the CVE program \citep{cvesite2024}. When a vulnerable package is used as a dependency, whether directly or indirectly, all dependent packages are exposed to the risk of inheriting that vulnerability. This mechanism of transitive vulnerability exposure is widely observed in empirical studies of {\tt npm} and other language ecosystems~\cite{decan2018evolution, zerouali2018empirical,jsdpendency}. 

\begin{figure}[th!]
\centering
    \includegraphics[width=0.4\textwidth]{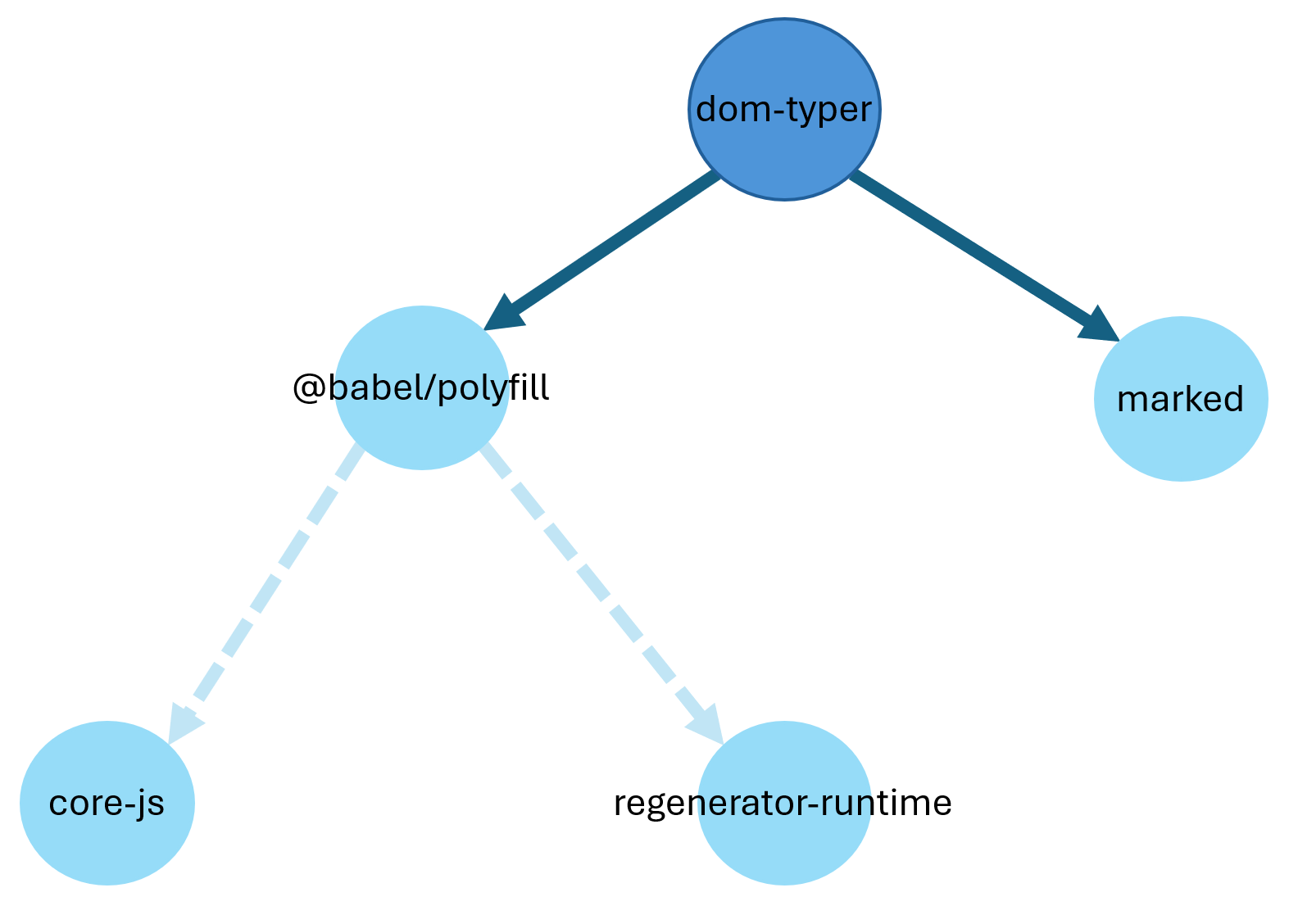}
    \caption{Example dependency graph for parent package `dom-typer', showing direct dependencies (dark blue) and transient ones (dashed, faded blue).}
    \label{fig:networkGraph}
    \vspace{-0.3cm}
\end{figure}

A \textit{direct dependency} refers to a package that is explicitly required by another package or project.  An \textit{indirect dependency} (or transitive dependency) is required by one of the direct dependencies, thus forming a multi-layered dependency chain.  Figure \ref{fig:networkGraph} demonstrates the dependency network for a given parent package and both direct and transient dependencies.  As can be seen, the parent package `dom-typer' is directly dependent on packages `@babel/polyfill' and `marked.'  Furthermore, the dependency, `@babel/polyfill', is dependent on packages `core-js' and `regenerator-runtime' - an indirect or `transient' dependency relative to the parent package `dom-typer'.  Both types of dependencies are critical in assessing vulnerability risks since vulnerabilities can propagate across these layers and throughout the dependency network.  This issue is particularly concerning when outdated dependencies serve as conduits for vulnerabilities--a common occurrence in software projects, especially JS-based. One study found that 60\% of all considered JS projects relied on outdated dependencies as of April 2018 \citep{zerouali2018empirical}.

Furthermore, vulnerabilities in the form of CVEs can vary in \textit{severity}, with their potential impact on a system measured if they were to be exploited. These severity levels are standardized into low, medium, high, or critical categories under the Common Vulnerability Scoring System (CVSS), developed by the FIRST group \citep{firstsite2024}. Identifying how these vulnerabilities propagate through dependency networks according to severity category is crucial for improving risk management and security response efforts.

Following established terminology from software supply–chain security research and formal frameworks such as CVE, CVSS, and NVD, we adopt definitions that align with prior empirical studies of package ecosystems \citep{decan2019empirical, zerouali2018empirical,zimmermann2019small}. These definitions enable consistent reasoning about vulnerability inheritance, dependency structure, and severity characterisation across 
{\tt npm}'s large-scale dependency network.

\noindent Formally, we state the following definitions:
\begin{mydef}
    \textbf{Vulnerability:} A weakness or flaw in a software package that can be exploited to compromise the confidentiality, integrity, or availability. A vulnerability becomes publicly known once assigned a CVE identifier and catalogued in NVD or OSV \citep{cvesite2024, osv2024}.
\end{mydef}

\begin{mydef}
    \noindent \textbf{Severity:} A standardised measure indicating the potential impact and exploitability of a vulnerability, expressed using some scoring frameworks such as the CVSS and categorised as Low, Medium, High, or Critical \citep{firstsite2024}.
\end{mydef}

\begin{mydef}
    \noindent \textbf{Package Dependency:} A software package that a given parent package relies on to function properly.  Typically a specific function/collection of functions of a package dependency is inherited.
\end{mydef}

\begin{mydef}
    \noindent \textbf{Package Dependent:} A package that relies on another package to operate correctly.
\end{mydef}
\noindent 
\begin{mydef}
    \noindent \textbf{Parent Package:} The primary or top-level package that declares dependencies on other packages.  Typically represents the project or module under analysis that brings in additional packages through its dependency declarations.
\end{mydef}
\noindent 
\begin{mydef}
     \textbf{Direct Dependency:} A package that is explicitly declared by the parent package as a dependency - it is directly required.
\end{mydef}

\begin{mydef}
    \noindent \textbf{Indirect/Transitive Dependency:} A package that is not directly declared by the parent package but is required by one of its dependencies.
\end{mydef}

\begin{mydef}
    \noindent \textbf{Outdated Dependency:} A dependency for which a newer stable version is available.
\end{mydef}

\subsection{Problem Statement}
In modern software ecosystems, packages often rely on one another to function correctly, forming complex networks of dependencies. Understanding the relationships between package dependencies and package dependents is crucial for assessing software security and performance. Identifying both direct and transitive dependencies, along with determining the packages affected by vulnerabilities in a given package, is essential for managing security risks, versioning, and maintenance.

This study aims to model and analyse the relationships between package dependencies and package dependents within software ecosystems. The objective is to develop an efficient method to enhance the understanding of systemic security risks in the {\tt npm} ecosystem. The findings will help software maintainers and developers prioritize security measures, improve vulnerability response times, and refine best practices for dependency management, ultimately strengthening the security and integrity of the ecosystem. Finally, we construct a vulnerability database that will serve as a valuable resource for the cybersecurity vulnerability detection research community.

\subsection{Differences With Prior Work}
Our work differs from prior studies in several key ways. As a baseline, we compare against the most relevant work by Decan et al.~\cite{decan2018impact}, who analysed 399 vulnerabilities across 269 npm packages over six years. First, their analysis considers only direct dependencies, while we model both direct and transitive dependencies to construct full package-level dependency networks and analyse vulnerability propagation. Second, their dataset was manually collected from Snyk.io, whereas we perform automated, large-scale aggregation from multiple authoritative sources (OSV.dev, deps.dev, NVD, and {\it npm-follow}), enabling broader coverage and cross-verification. Third, instead of static CSV summaries, our structured database links 27 features across packages, CVEs, CWEs, dependencies, and temporal metadata, supporting predictive modeling, risk scoring, and multi-dimensional analysis. Finally, while their scope ends pre-2017, our dataset scales to over one million npm packages, providing a modern and comprehensive 2024 ecosystem view.

Unlike Liu et al.~\cite{liu2022demystifying}, who do not consider vulnerability severity and rely on manual dependency-tree resolution, our fully automated and reproducible pipeline incorporates historical publication, modification, and fix dates and applies quantitative CVSS v3.x metrics (base, exploitability, impact) to support consistent severity assessment and vulnerability-risk propagation. Zimmermann et al.~\cite{zimmermann2019small} focus on descriptive ecosystem characteristics--package influence, maintainer roles, and limited known vulnerabilities -- without constructing full dependency networks or integrating multi-source vulnerability data. In contrast, our work enables network-level, temporal, and predictive vulnerability analysis using a large-scale, structured, and multi-source dataset.

\section{Methodology}
\label{sec:methodology}

\begin{figure}[h]
    \begin{center}
        \includegraphics[width=0.4\textwidth]{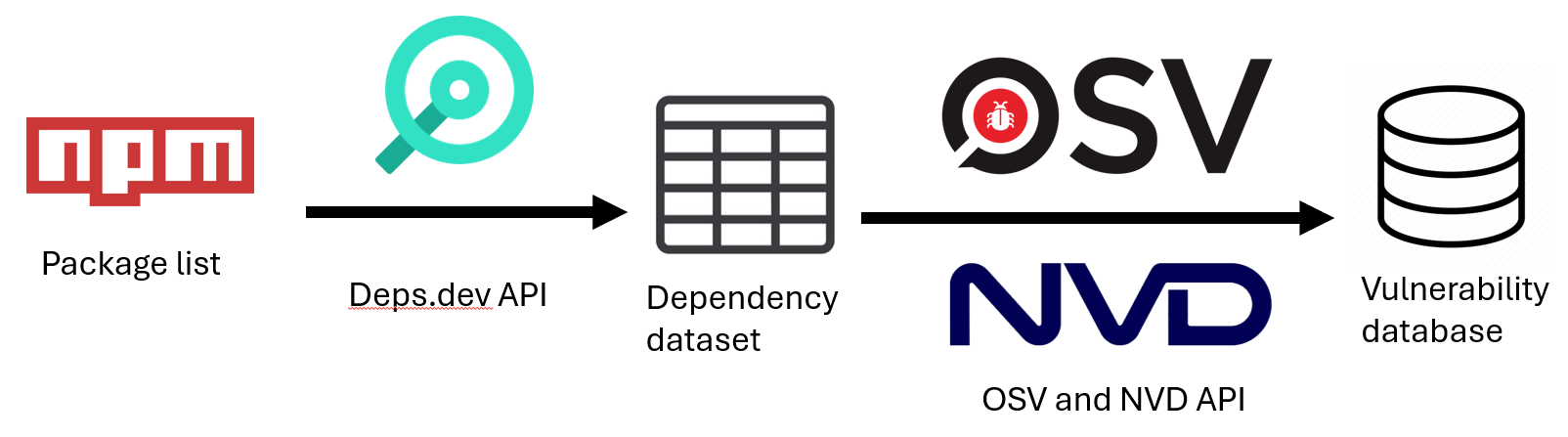}
        \end{center}
        \caption{Pipeline describing our methodology for the creation of vulnerability database.}
        \label{fig:methodOverview}
    \end{figure}

In our research, we develop a dataset of dependency networks and a vulnerability database to examine software security risks. As illustrated in Figure \ref{fig:methodOverview}, our process starts with extracting dependency networks from a specified list of packages, mapping their interconnections to interpret how vulnerabilities spread. For each package, we focus on the default or most recent stable version to maintain relevance. We then detect vulnerabilities using the Google Open Source Vulnerability (OSV) API and the National Vulnerability Database (NVD) API, which offer CVE identifiers, CVSS severity scores, and descriptions of vulnerabilities. This information aids in evaluating security risks across software ecosystems. The gathered data is organized into a structured vulnerability database, listing package names, versions, severity levels, and advisory references. This database supports a thorough risk assessment by pinpointing security flaws within dependency networks.

Furthermore, by keeping a historical record of vulnerabilities, we enable longitudinal analysis, allowing researchers to monitor security trends over time. 
In this paper, our study offers a systematic approach to vulnerability analysis within software dependencies, incorporating real-time security intelligence to improve risk assessment and mitigation strategies in contemporary software ecosystems. The following subsections provide a detailed overview of vulnerability database creation, including platform, data sources, data linkage, and analysis.

\vspace{-2mm}\subsection{Platform \& Data Sources}
\label{subsec:dsource}
For the past decade, JS has remained one of the most popular languages on GitHub by usage, with approximately 6.7M projects as of October 2024 \citep{daigle2023octoverse}.  As the main package manager for the JS language, the {\tt npm} also hosts approximately 4 million unique open-source packages and additionally possesses larger and more complex dependency networks than other package distributions \citep{decan2019empirical}.  Due to the {\tt npm}'s popularity, we expect a large community of contributors necessary to identify and publish vulnerabilities found in JS packages and projects on various vulnerability databases.  As such, due to its popularity and degree of established complexity, we have chosen to investigate the presence of vulnerabilities throughout dependency networks within the {\tt npm} ecosystem.

The packages gathered for our dataset were attained through the {\tt `npm-follower'} project, which provides weekly snapshots of all available packages on the {\tt npm}, including both source code and metadata information \citep{pinckney2023npm}.  As of 5 September 2023, the {\tt npm} ecosystem comprised 3,946,361 unique packages and 36,432,989 unique package versions according to the dataset attained. From this extensive dataset, we randomly selected 1.5 million unique JS package names. Sampling was performed without replacement using Python's {\tt random.sample} method to ensure uniqueness.  Post-sampling validation removed approximately 400,000 entries due to unavailability in the live repository or designation as deprecated/malicious package stubs.  This filtering resulted in a subset consisting of 1,077,946 {\tt npm} parent packages (approximately a quarter of all available packages)-- a large enough proportion to provide a representative sample of the {\tt npm} ecosystem and maintain a manageable data size for in-depth analysis.

To analyse the presence of vulnerabilities throughout dependency networks, we must first construct the individual dependency networks for each parent package within our gathered dataset.  For this, we are extracting dependency information through the use of the Open Source Insights {\tt deps.dev} database \citep{depsdev2024}--a service developed and hosted by Google that constructs a full graph of OSS package dependency networks, including dependency package names and the latest required versions.  For parsing, we usually extract the default version as stored on the {\tt npm} registry--if there is no default version, we extract the latest stable release version of the package to have its given dependency network constructed.  Table \ref{tab:data_overview_pkglist} summarises our dataset from our package list.

\begin{table}[ht!]
    \centering
    \caption{Overview of the analysed package dataset and its features.}
    \begin{tabular}{p{2cm}|p{5.5cm}}
    \toprule 
    Packages & 1,077,946 \\ \hline 
    Total Features & 11 \\ \hline
       Feature Names  & pkg\_name,	default\_ver,	direct\_dependencies,	indirect\_dependencies,	dependencies\_count,	dependencies\_names,	dependency\_graph,	cve\_count,	cve\_labels,	cwe\_count,	cwe\_labels \\

       \bottomrule 
        
    \end{tabular}
    
    \label{tab:data_overview_pkglist}
\end{table}

The core of our vulnerability database is developed around the dependency networks of these {\tt npm} parent packages.  Utilizing the constructed dependency networks dataset, we identify known vulnerabilities associated with relevant dependencies and associated version numbers using the NVD and Google OSV APIs \citep{nvd2024, osv2024}.  Both APIs are queried to find the earliest date the vulnerability was published to each respective database, as discrepancies can exist between these dates. 
Furthermore, we store any relevant information that may assist in discerning relevant insights, such as severity levels, fix dates, affected packages, and more.  The final output is a database containing each vulnerability found within our collection of dependency networks to be queried as relevant for our purposes.  Table \ref{tab:data_overview_vulns} presents a brief overview of the vulnerability database.

\begin{table}[ht!]
    \centering
    \caption{Vulnerability database and its features.}
    \begin{tabular}{p{2cm}|p{5.5cm}}
    \toprule 
    Vulnerabilities & 1,517 \\ \hline 
    Total Features & 27 \\ \hline
       Feature Names  & cve\_label,	osv\_label,	summary,	details,	registered\_package,	affected\_packages, dependents,	osv\_published, nvd\_published, earliest\_publish,	osv\_modified, introduced\_ver, introduced\_date, fixed\_ver, fixed\_state, last\_affected\_ver, lastVuln\_date, limit,	database\_specific,	references,	affected, schema\_version, severity, baseSeverity, baseScore, exploitabilityScore, impactScore \\

       \bottomrule
        
    \end{tabular}
    
    \label{tab:data_overview_vulns}
    \vspace{-0.3cm}
\end{table}

\subsection{Data Linkage}
\label{sec:vuldatalinkage}
Data linkage in software ecosystems involves integrating information from multiple sources to form a unified dataset. Specifically, this process links package metadata from databases or repositories such as {\tt npm} with vulnerability records from databases like the NVD. This linkage enables a comprehensive analysis of vulnerabilities across ecosystems. In Appendix \ref{appendix:algo}, we provide details about the algorithm we developed to construct the unified database.

\subsection{Data Analysis}

In investigating the proposed research questions, we implement a variety of statistical and plotting tools to convey our insights coherently.  For $\mathbf{RQ_1}$ and $\mathbf{RQ_2}$, when investigating the duration spanning certain key dates, we take advantage of the statistical technique of survival analysis--similar to other works \cite{zerouali2022impact, alfadel2023empirical, decan2018impact}.  Survival analysis focuses on predicting the time until an event occurs, such as failure or death, and is notable for handling data where the given event has not happened for some subjects by the study's end (referred to as right-censored data).  Our case focuses on whether a vulnerability still exists without a fix and, if fixed, how long it persisted.  

The Kaplan-Meier estimator, a cornerstone of survival analysis, is a non-parametric method used to estimate the probability of surviving past certain times based on observed event times.  To elaborate, the estimator of the survival function $S(t)$ is defined as the product of conditional survival probabilities at observed event times and is given by:

\begin{equation*}
    \hat{S}(t)=\prod_{i: t_i \leq t}(1- \frac{d_i}{n_i})
\end{equation*}

where $t_i$ denotes the ordered time points at which the events occur, $d_i$ is the number of events at time $t_i$, and $n_i$ is the number of subjects yet to expire just prior to $t_i$.  This step function, hence, provides an estimate of the probability that a subject survives beyond time $t$, effectively incorporating right-censored data without assuming a specific distribution for survival times.  In our case, it may be used to determine the proportion of fixed vulnerabilities starting from the point at which they are found to exist, for example.

The median survival time in a Kaplan–Meier distribution may be obtained by identifying the time at which the estimated survival function $S(t)$ first drops to 50\%, representing the point at which half the population is expected to have experienced the event. Because the median depends only on reaching the 50\% survival threshold—and the Kaplan–Meier estimator explicitly accounts for right-censored observations—it is less influenced by extreme or heavily censored values, thereby reducing the skewing effects seen in mean-based measures.

Furthermore, we use regression analysis to estimate the trend of vulnerabilities over time for $RQ_1$ and the correlational relationship of dependencies and vulnerabilities in $RQ_3$, considering both linear and exponential models. To identify the most appropriate model for these relationships, we use the $R^2$-value, a statistical measure that indicates the proportion of variance in the dependent variable explained by the independent variable. In simple terms, the $R^2$-value reflects the goodness-of-fit of a model: values closer to 1 suggest a better fit, while values closer to 0 indicate a poorer fit.

\section{Empirical Study}
\label{sec:study}
Our study investigates the lifecycle of vulnerabilities in the {\tt npm} ecosystem, focusing on how vulnerabilities are published, fixed, and propagated through dependency networks. Prior work has largely examined mean disclosure delays or small-scale case studies; however, the realities of modern software ecosystems--highly interconnected, versioned, and long-lived--require a more robust temporal model. Therefore, we focus our analysis on survival-curve–based \textit{medians} and align all timelines with the first affected release, capturing the full exposure window experienced by downstream users. With this framework, we examine three research questions (mentioned in \S~\ref{sec:intro}) in the following subsections. 

\subsection{\texorpdfstring{$\mathbf{RQ_1}$}: What is the nature of vulnerability discovery and publication on the {\tt npm} ecosystem?}
\label{subsec:rq1}

Before presenting our results, we clarify our temporal methodology, as it underpins both $\mathbf{RQ_1}$ and $\mathbf{RQ_2}$. All timeline measurements in this study are derived from Kaplan–Meier survival curves, which allow us to model right-censored vulnerabilities and report \textit{medians} rather than arithmetic means. This is essential because fix and publication delays exhibit heavy-tailed distributions that cannot be represented meaningfully using simple averages.

A second methodological distinction is our alignment of timelines to the first affected release—the date the vulnerable version was uploaded to {\tt npm}. Prior ecosystem studies typically align with advisory publication (e.g., NVD or OSV timestamps), which omits long periods during which users were already exposed. By anchoring at the moment of introduction, we capture the true exposure interval and enable an ecosystem-wide analysis of the ordering between publication and remediation.

As shown later in Figure~\ref{fig:fixFromPub}, this alignment reveals that 83.1\% of vulnerabilities are fixed before they are publicly published, with a median publication-fix offset of $-19$ days. This ordering result, to our knowledge, has not been documented at scale for {\tt npm} and provides new insight into how disclosure and remediation interact in practice.
With this methodological foundation, in $RQ_1$, we now provide developers a sense of whether vulnerabilities are increasing at a manageable rate and further suggest whether developers are publishing vulnerabilities in their projects within a reasonable time frame relative to the first known exposed version.  By first understanding some basic summary statistics and the trend of vulnerabilities over time, we may indicate whether greater awareness of vulnerabilities in packages and their propagation through dependency networks is required.

\begin{table}[h!]
\centering
\caption{Summary statistics of the vulnerability database.}
\begin{tabular}{r | l}
\toprule
Total {\tt npm} parent packages: & 1,077,946 \\
Parents dependent on a vulnerable package: & 232,836 \\
\hline
Total unique dependencies: & 421,941 \\
Total releases of dependencies: & 658,136 \\
\hline
Total unique {\tt npm} packages: & 1,363,977 \\
Total unique releases: & 1,619,856 \\
\hline
Total vulnerability instances: & 2,150,104 \\
Unique vulnerabilities (All CVSS versions): & 1,517 \\
Unique vulnerabilities (CVSS v3.x): & 1,498 \\
Affected packages: & 1,083 \\
\bottomrule
\end{tabular}
\label{table:summaryStats}
\end{table}

\begin{figure}[ht!]
    \centering
    \includegraphics[width=0.35\textwidth]{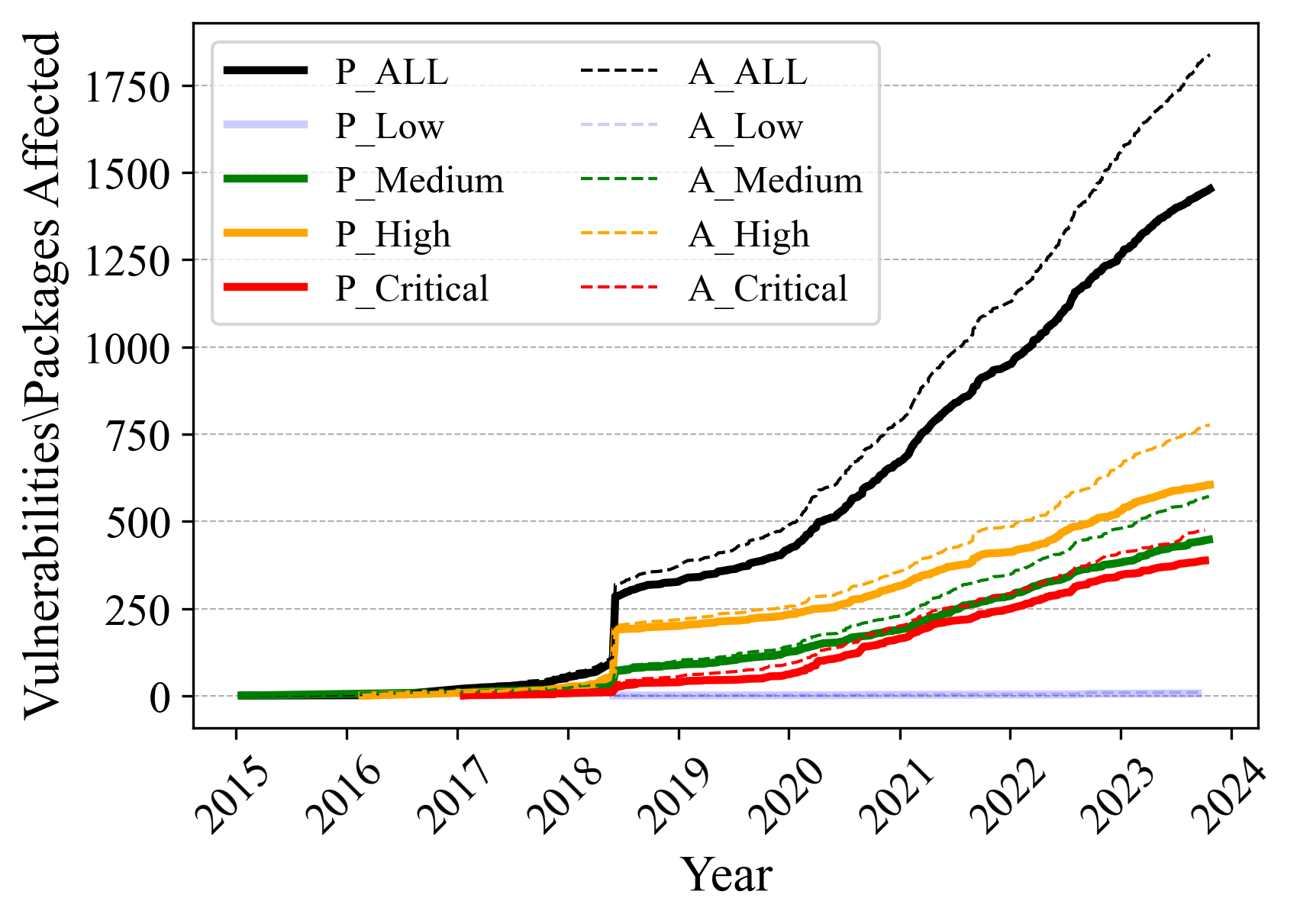}
    \caption{Accumulation of the number of published vulnerabilities (solid lines: Prefix P\_) and corresponding affected packages (dashed lines: Prefix A\_). 
    }
\label{fig:Fig1}
\end{figure}

As summarized in Table \ref{table:summaryStats}, of the 1,077,946 parent packages in our database, 21.60\% (232,836) had at least one vulnerability in their dependencies.  There are 421,941 unique dependencies, of which 658,136 unique versions of these dependencies exist.  In total, including parent packages and dependencies, there exist 1,363,977 unique packages across the entire database.  Additionally, there exist 2,150,104 instances of vulnerabilities throughout the entire database, belonging to 1,515 unique CVE types.  Excluding those that are under the CVSS \texttt{v2.0} scheme, 1,498 vulnerabilities exist, consistent with CVSS \texttt{v3.x}.
Figure \ref{fig:Fig1} shows the accumulation of {\tt npm} vulnerabilities published to the OSV database, whereby solid lines refer to accumulated vulnerabilities while dashed lines refer to the accumulation of affected packages.  The trend of Figure \ref{fig:Fig1} seems to straightforwardly suggest that the number of vulnerabilities published is increasing over time, with the number of affected packages increasing at a higher rate for all severity categories.  This seems reasonable as each vulnerability can affect one or more distinct packages.

An interesting feature of this plot is the large influx seen around June 2018.  By investigating the associated vulnerabilities published within the given period, this large increase seems to have been done across 3 weeks by the HackerOne group \footnote{For example, see the `Change History' section of CVE-2017-16138 at \url{https://nvd.nist.gov/vuln/detail/CVE-2017-16138}.}.  Additionally, since these vulnerabilities belong to CVSS \texttt{v3.0} and \texttt{v3.1} exclusively, it may be that there was a large increase in the use of this convention in mid-2018, as opposed to CVSS \texttt{v2}--though this remains speculative.  It must also be noted that it is very likely that many vulnerabilities are yet to be discovered. However, this increasing trend may suggest an increase in developers' ability to detect vulnerabilities through more robust tools and effective discovery techniques.  The relative `flattening' of the cumulative line nearing the end of the timeline may be because it takes around 1 year, on average, for a vulnerability to be published relative to its discovery date, as found in prior research \cite{decan2018impact}.

Considering vulnerabilities from all categories, accumulation seems to occur linearly--as proven statistically using regression analysis of a linear and exponential growth model.  The coefficient of determination, $R^2$, suggesting the goodness-of-fit for both models, is displayed in Table \ref{table:reg_values}.  This linear increase is also characteristic of an increase in affected packages.  These findings suggest that while threats are steadily increasing, the growth is not trending at an undesirable rate, which might allow for more measured budgeting and resource allocation approaches.  However, this also suggests that the rate of discovery has not increased in any significant manner, implying that efforts may not be as concentrated as required.

\begin{table}[hb]
\centering
\caption{$\mathbf{R^2}$-values for regression analysis on accumulating vulnerabilities and affected packages over time. }
\begin{tabular}{l|c|c}
\toprule

\multicolumn{1}{c|}{\textbf{Model}} & \textbf{No. Vulnerabilities} & \textbf{No. Packages} \\
\hline
Linear & \underline{0.97} & \underline{0.96} \\
Exponential & 0.69 & 0.69 \\
\bottomrule
\end{tabular}
\label{table:reg_values}
\end{table}

Table \ref{tab:vul_score_category} provides a clearer representation of the final distribution of severity categories within the database. We find that out of 1,498 unique vulnerabilities, a large majority are either medium, high, or critical in severity (466, 630, and 394, respectively), whereas there are only 8 vulnerabilities of low severity.  The lack of low severity vulnerabilities could be due to a reporting bias that increases the potential risk of assessed vulnerabilities \citep{spring2021time} or the potential that maintainers do not consider low-risk vulnerabilities worth disclosing.  The amount of vulnerabilities belonging to the critical severity category seems alarming, especially due to its constrained score range (9.0 to 10.0), though this may be inflated as severity scores exceeding 10.0 are reduced to a limit of 10.0, as per the CVSS scheme \citep{nvd2023}. We find that less than 1\% of vulnerabilities belong to the `Low' category, 31\% belong to the `Medium' category, a large portion (42\%) are classified as `High' severity, and the remaining 26\% are found to be of `Critical' severity.

\begin{table}[ht!]
    \centering
    \vspace{-0.25cm}
     \caption{Distribution of 1,498 vulnerabilities for each severity category.}
     \vspace{-0.25cm}
    \begin{tabular}{ c c c c}
    \toprule
     \textbf{Score} & \textbf{Category} & \textbf{Number} & \textbf{Percentage}\\ \hline 
      $<$ 4.0 & Low & 8 & 0.53\% \\
       $\geq$ 4.0 and $<$ 7.0 & Medium & 466 & 31.11\% \\
       $\geq$ 7.0 and $<$ 9.0 & High & 630 & 42.06\%\\
       $\geq$ 9.0 & Critical & 394 & 26.30\%\\ \bottomrule 
    \end{tabular}
    \label{tab:vul_score_category}
\end{table} 

\begin{figure}[htp]
    \centering
     \includegraphics[width=0.98\linewidth]{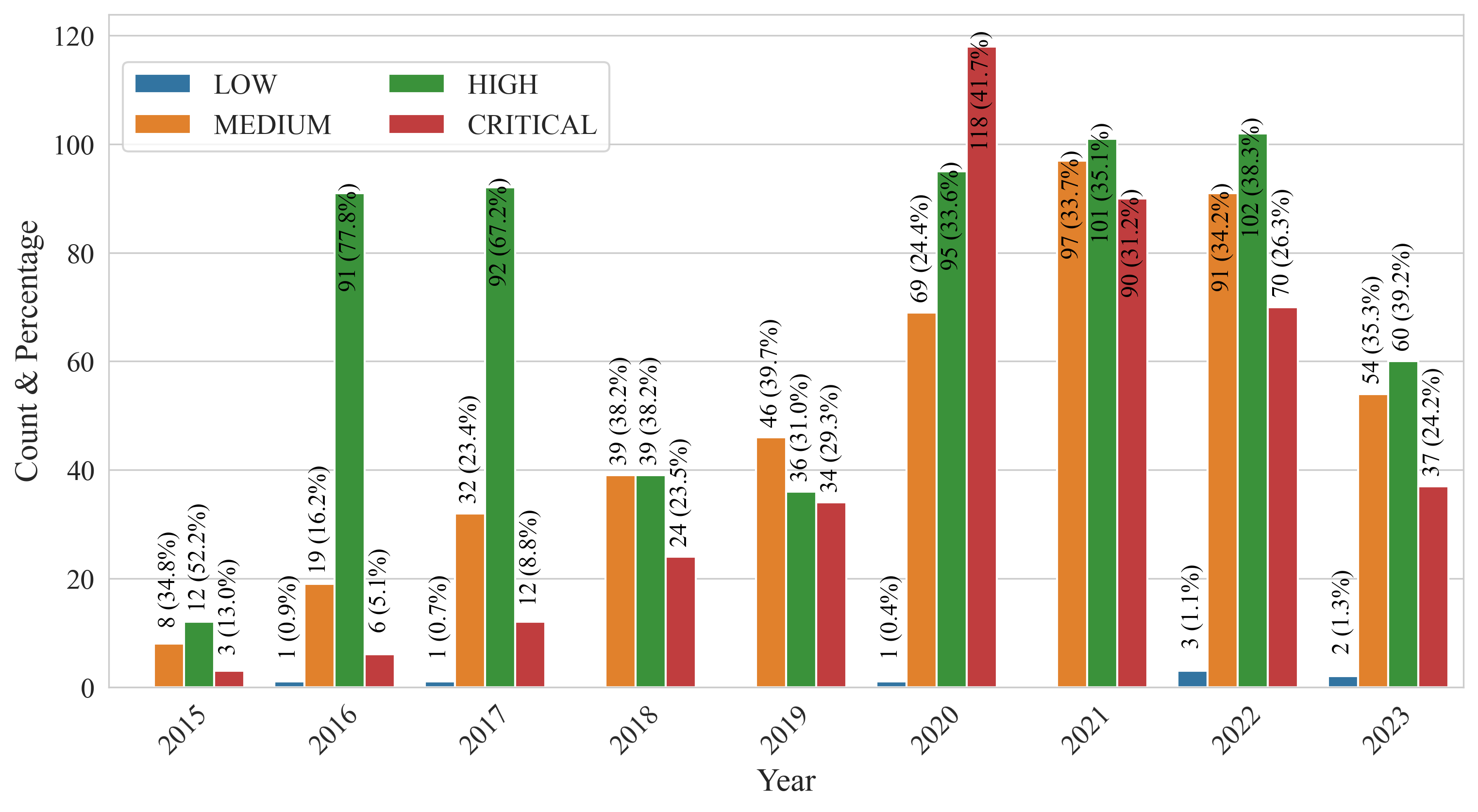}
     \vspace{-0.15cm}
    \caption{Trend of various (LOW, MEDIUM, HIGH and CRITICAL) vulnerabilities over the years. } 
    \label{fig:vul_trend}
    \vspace{-0.3cm}
\end{figure}

Figure \ref{fig:vul_trend} illustrates the distribution of various vulnerabilities and their year based percentage from 2015 to 2023. The number of critical vulnerabilities increased steadily from 2015, peaking at 118 in 2020. Subsequently, a downward trend was observed, with the count declining to 37 by 2023. This trend suggests that developers have prioritized addressing critical vulnerabilities, likely due to their significant impact on system security. A similar trend is observed for medium-severity vulnerabilities, which initially increase, peaking at 97 in 2021, before declining thereafter. The trend for high-severity vulnerabilities follows a different pattern but exhibits a decline from 2022 to 2023. Overall, when considering all vulnerabilities per year, the total number in 2023 is lower compared to the preceding years (2020–2022). This suggests that developers have become more proactive in addressing vulnerabilities before releasing their software.
The percentage of critical vulnerabilities peaked at 41.70\% in 2020 but gradually reduced to 24.18\% by 2023. Throughout most years, high-severity vulnerabilities accounted for the largest share, except in 2019. This trend indicates that developers prioritize addressing critical vulnerabilities more swiftly compared to other types.

\begin{figure}[ht!]
    \centering
    \includegraphics[width=0.9\linewidth]{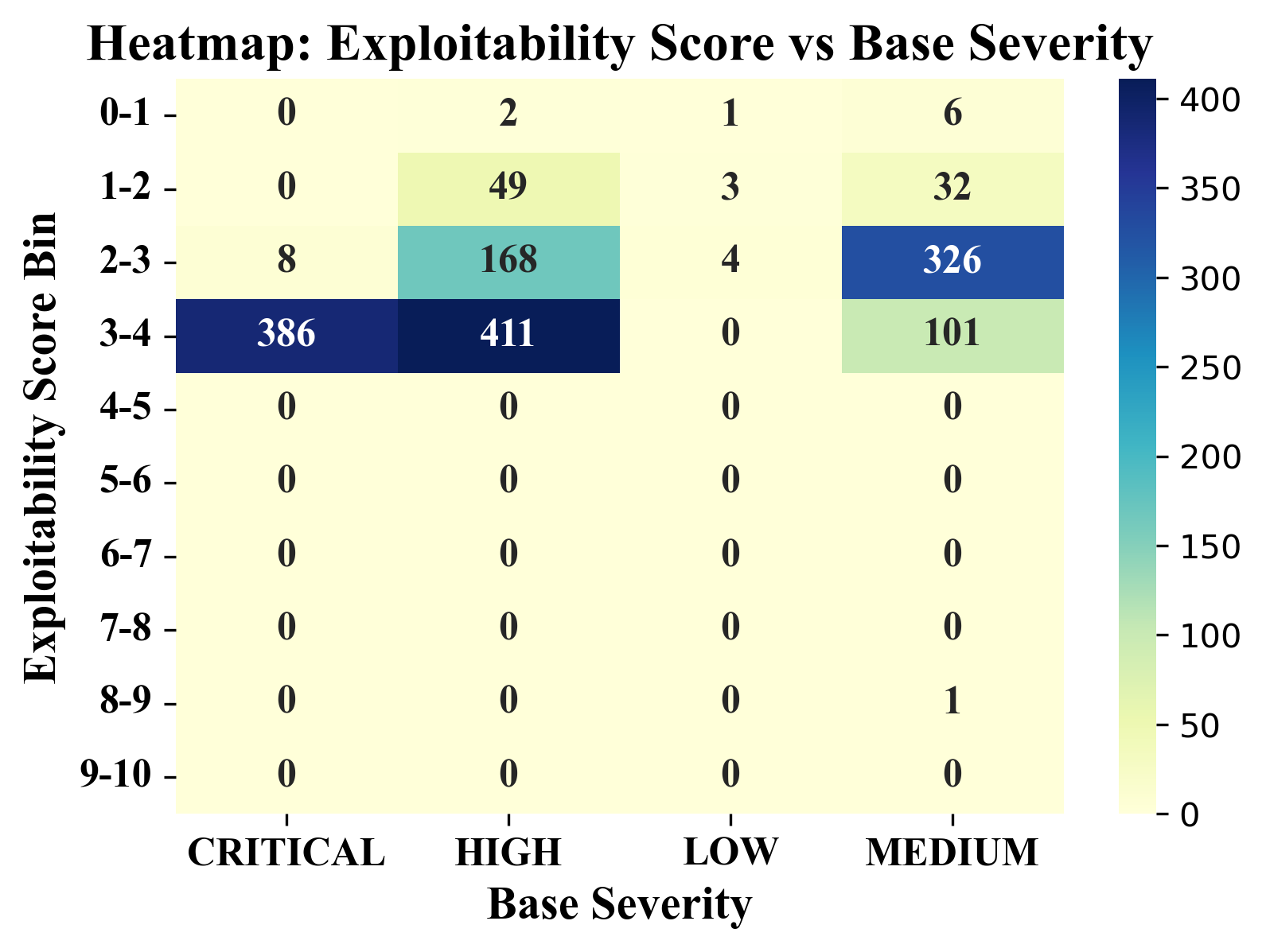}
    \vspace{-0.15cm}
    \caption{Heatmap of Base Severity and Exploitability Score.}
    \label{fig:heatmap}
\end{figure}

Severity classification may be important; however, exposure within a network is more crucial to assessing vulnerability risk, as it may elucidate how readily a flaw can be turned into an attack. Though severity scores are determined as a combination of impact and exploitability metrics, we capture the exploitability scores specifically, as they are vital for fix prioritization.  Figure \ref{fig:heatmap} shows the relationship between vulnerability severity and exploitability -- demonstrating a strong correlation between `Critical' and `High' severity vulnerabilities and moderately high exploitability scores ranging from values of 2 to 4, indicating that many of the most dangerous flaws are also relatively easy to exploit. This highlights the urgent need to address these vulnerabilities. Conversely, `Low' severity vulnerabilities generally have lower exploitability. This figure motivates a prioritization strategy that considers a combination of severity and exploitability, enabling security teams to optimize resource allocation and effectively mitigate the most pressing risks.

We assess how promptly developers disclose vulnerabilities by measuring the median time between the release of the first vulnerable npm version and its public publication in NVD or OSV. The duration between these events could help reveal whether publication times are reasonable and whether this may have subsequent contributions to delayed or expedited fix-dates--for example, if publication takes a long time, then the fix-date could correspondingly appear long.

\begin{figure}[h!]
\centering
    \includegraphics[width=0.4\textwidth]{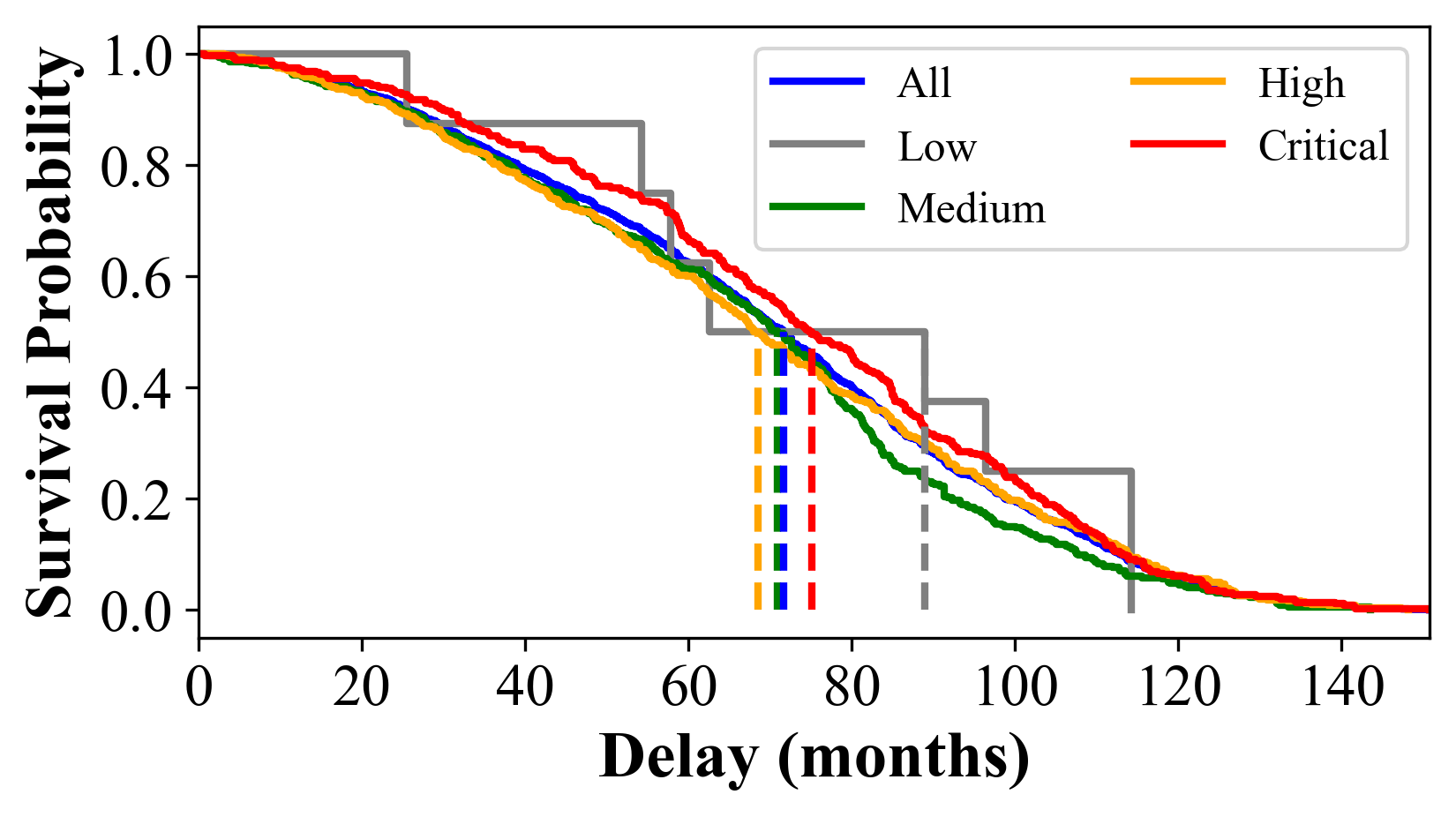}
    \caption{Kaplan-Meier distribution for event ``vulnerability is published'' concerning the date of first affected release.}
    \vspace{-0.25cm}
    \label{fig:pubFromInitialCreation}
\end{figure}

Figure \ref{fig:pubFromInitialCreation} presents the survival curves measuring the time from when a vulnerable version is first released in npm to when the corresponding vulnerability is published in OSV or NVD. Across all severity levels, the median time to publication is extremely long—approximately 71 months (5 years and 11 months)—indicating that disclosure substantially lags the initial introduction of the vulnerability. The curves for high, medium, and critical severities follow similar trajectories, showing little variation in publication timelines. Although critical vulnerabilities appear to take slightly longer to reach public advisories, this difference is small relative to the overall timescale. The low-severity curve shows greater variability but is based on a comparatively small sample and should be interpreted cautiously. Overall, vulnerabilities remain undisclosed for long periods, highlighting the need for more timely detection and reporting. 
The close alignment of the survival curves across severity categories suggests that the ecosystem's disclosure process does not meaningfully differentiate based on severity. This may reflect workflows in which publication timing is influenced more by reporting practices, identification delays, and ecosystem processes than by the underlying criticality of the vulnerability. Although {\tt npm}'s coordinated disclosure policy enforces a 45-day embargo window before publication~\citep{npm2024}, this mechanism is negligible relative to the multi-year median disclosure lag observed here. Overall, these results indicate systemic delays in vulnerability publication that affect all severity levels in a similar manner.

\noindent\fbox{\parbox{0.98\linewidth}{
\textbf{{\textit Findings.}} Vulnerabilities seem to be increasing linearly, with the number of affected packages increasing at a higher rate.  Most of the vulnerabilities belong to the high and medium severity categories, with a substantial amount belonging to the critical category.  Out of the 1,077,946 parent packages, 21.60\% (232,836 parent packages) contain at least one vulnerability throughout their dependency network.  The majority (42\%) of these vulnerabilities belong to the `High' severity category. However, this is closely followed by that of the `Medium' (31\%) and `Critical' (26\%) categories as well.  Furthermore, it takes approximately 5 years and 11 months for 50\% of vulnerabilities to be published onto Google OSV or the NVD from when the first vulnerable version is uploaded to the {\tt npm}.  Vulnerabilities of `Critical' severity seem to take approximately 4 months longer to be published. }} 


\subsection{\texorpdfstring{$\mathbf{RQ_2}$}: How effective is the response to discovered vulnerabilities on the {\tt npm} ecosystem?}

To understand the effectiveness of the response to discovered vulnerabilities on the {\tt npm} ecosystem, we need to understand how long packages may be vulnerable and potentially expose dependent packages to a given vulnerability.  As such, $RQ_2$ focuses on whether the duration packages are being affected by vulnerabilities and how soon after publication a fix is released.

\begin{figure}[th]
\centering
    \includegraphics[width=0.4\textwidth]{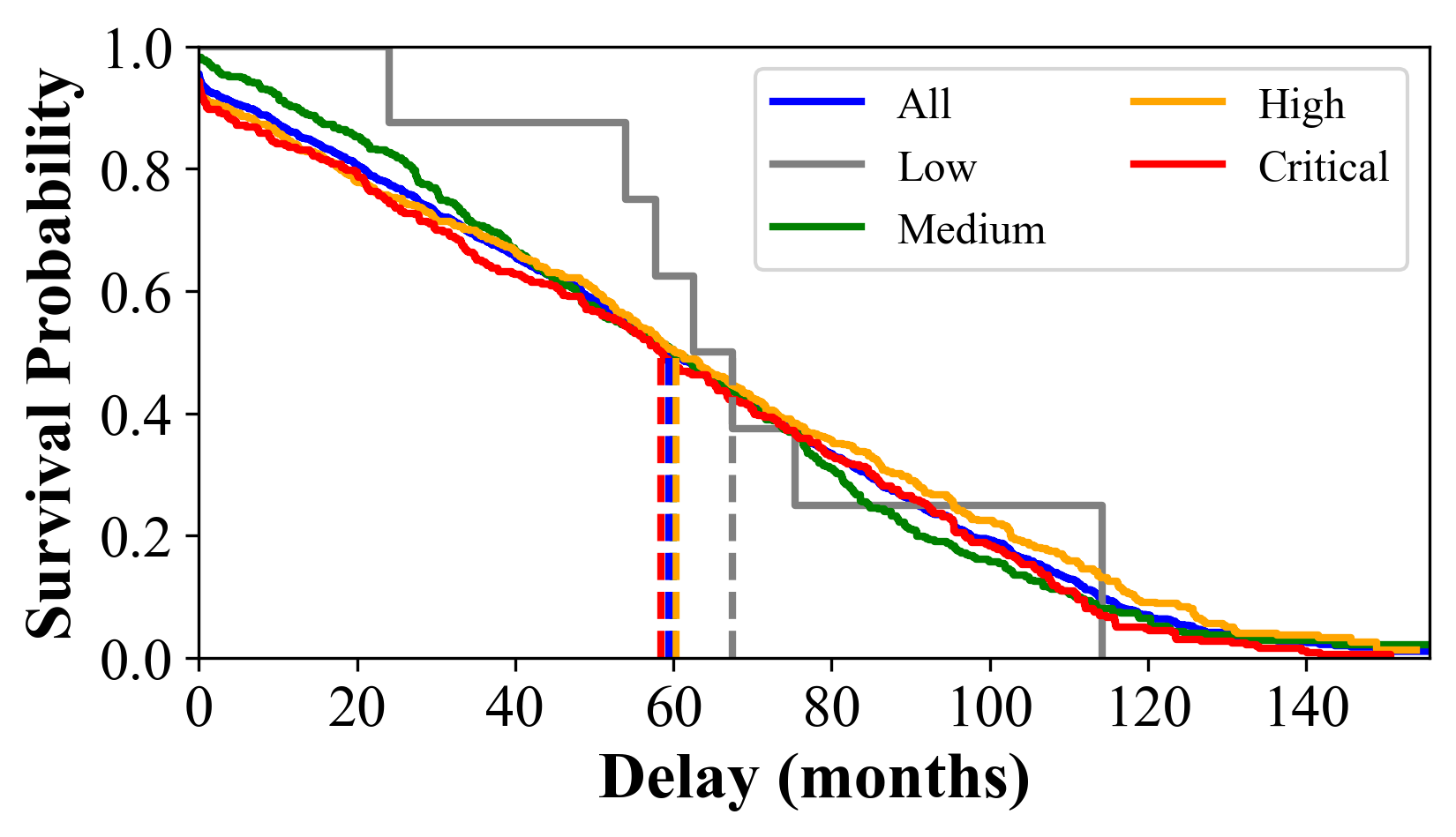}
    \vspace{-0.15cm}
    \caption{Kaplan-Meier survival distribution for event ``vulnerability is fixed'' with respect to the date of first affected release.}
    \vspace{-0.15cm}
    \label{fig:fixFromInitialCreation}
\end{figure}

Figure \ref{fig:fixFromInitialCreation} shows the Kaplan-Meier survival distributions for the duration from the first affected release of a package to the ``fix'' event--whereby a fix is available for the vulnerable package.  It seems, regardless of the severity category, it takes a rather long time for a vulnerability to be fixed--approximately 59 months (or 4 years and 11 months) for 50\% of all vulnerabilities to be fixed, which is common among all categories, similar to results for $RQ_1$.  One interesting feature regarding the previous section results is that the median duration to fix a vulnerability is around 1 year shorter than the median duration to publish, implying that vulnerabilities may be fixed before publication, on average.  To further investigate this phenomenon, we now determine the time taken for a vulnerability to be fixed relative to the publication date.  The intuitive result would be that the fix will be released at a date soon after publication, though most are being delayed due to development time.

\begin{figure}[ht!]
    \centering
       \includegraphics[width=0.4\textwidth]{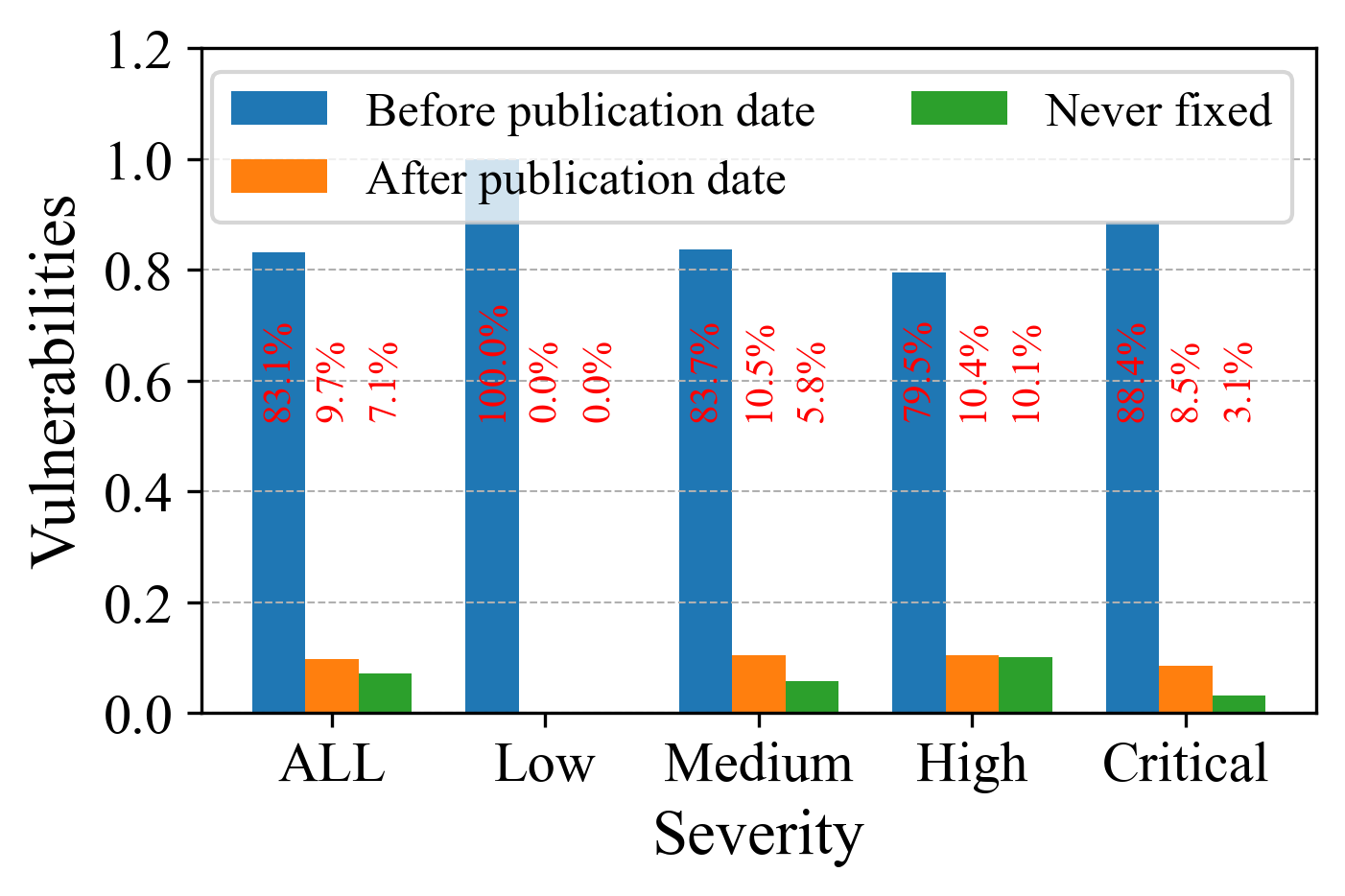} 
       \vspace{-0.15cm}
       \caption{Distribution of vulnerabilities by severity at the time of remediation. } 
        \label{fig:fixMoments}
        \vspace{-0.3cm}
    \end{figure}

When investigating the duration from publication to the fix-date, we found a large majority of instances possessed negative durations - meaning the vulnerabilities were being fixed before their publication.  As can be seen in Figure \ref{fig:fixMoments}, around 83.1\% of all vulnerabilities have their fix available \textit{before} publication date.  Whereas around 9.7\% more intuitively have their fix ready after the publication date, and the remaining 7.1\% have no fix available yet.  All categories of severity seem to mimic this distribution proportionally, with a large majority being fixed before publication and a minority being fixed after publication or never fixed at all.

The high frequency of vulnerabilities being fixed before their publication date can be attributed to several key factors. First, package maintainers often release updated versions that address security issues without explicitly identifying them as vulnerability fixes. Although this proactive maintenance helps protect users, it commonly involves major version updates that can introduce compatibility challenges. To minimize disruption, further analysis should examine whether these pre-publication fixes could be implemented through minor updates or patches instead.  Another significant factor is the {\tt npm} security protocol, which enables developers to address known vulnerabilities before they are publicly disclosed. This approach allows development teams to implement fixes preemptively, reducing the window of opportunity for malicious actors to exploit published vulnerabilities. By resolving security issues before publication, developers can better protect their users while maintaining the integrity of their dependencies.

\begin{figure}[ht!]
\centering
    \includegraphics[width=0.4\textwidth]{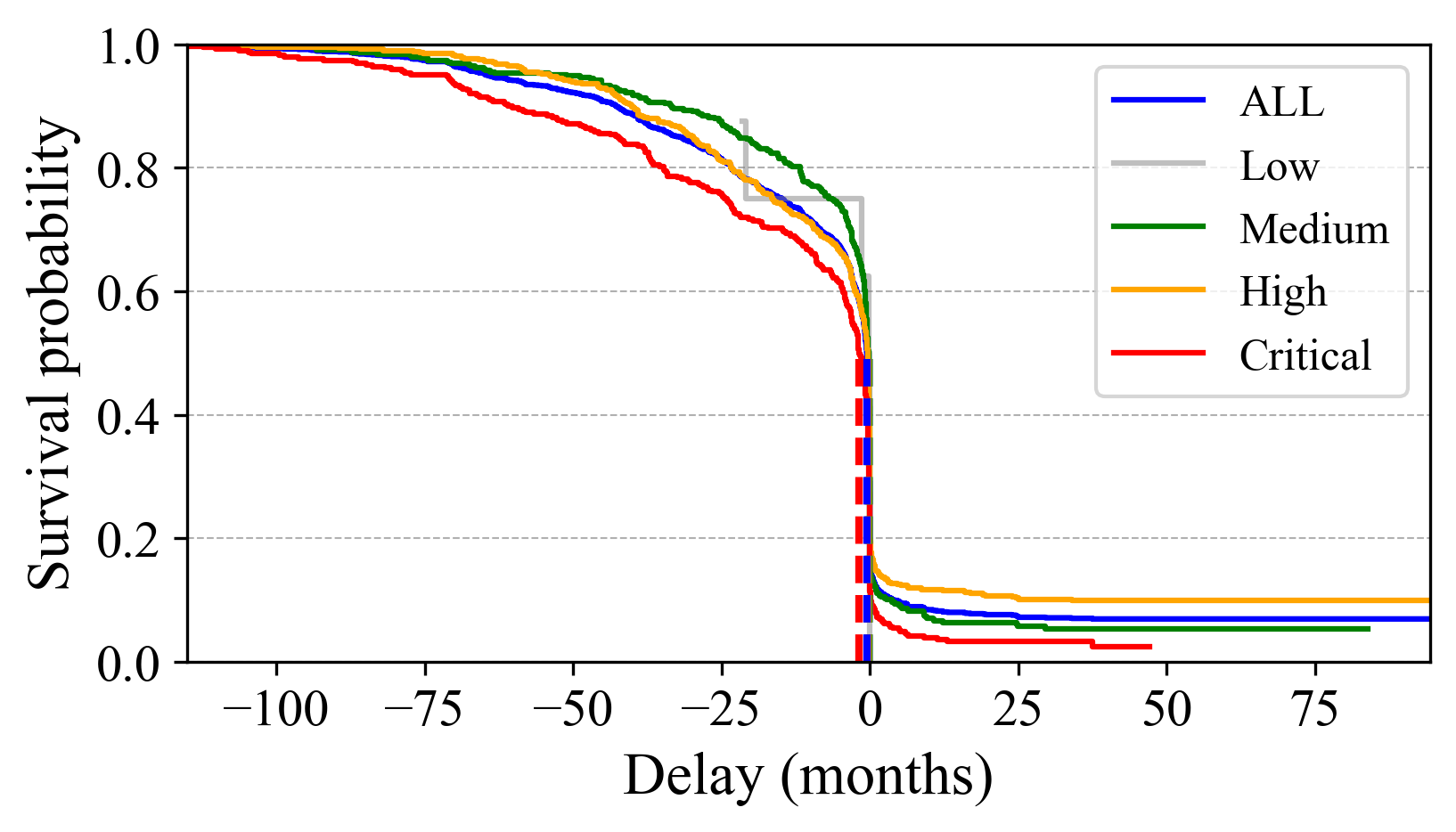}
    \caption{Survival probability for event ``vulnerability is fixed'' with respect to the vulnerability publication time. 
    }
    \label{fig:fixFromPub}
    \vspace{-0.3cm}
\end{figure}

Figure \ref{fig:fixFromPub} presents the Kaplan-Meier distribution of the durations from the vulnerability publication date to the fix-date.  As can be seen, the median duration at which 50\% of all packages published have been fixed is approximately -0.62 months (-19 days) - again, reaffirming the results prior.  Since most of these vulnerabilities seem to be fixed around the zero duration mark, it could be implied that most developers are publishing the vulnerability as soon as they have a fix developed--otherwise, the median duration would be noticeably positive.  This, again, reaffirms the potential motivation of developers to update their vulnerable packages before publication to mitigate the risk of malicious actors taking advantage of a vulnerability if it were published without a fix.
\noindent\fbox{\parbox{0.99\linewidth}{
\textbf{{\textit Findings.}} It takes approximately 4 years and 11 months to fix a vulnerable package, on average, from when the first vulnerable version is uploaded to the {\tt npm}.  There is also a lack of discrepancy between different severity categories when considering the duration of a corresponding fix.  Additionally, it takes approximately -19 days for 50\% of vulnerabilities to be fixed from their publication date on the NVD, as developers tend to publish after the fix has been released.  Accordingly, 83.4\% of all vulnerabilities are fixed before their publication date, 9.7\% are fixed after their publication date, and 6.9\% have no fix available yet. }} 


\subsection{\texorpdfstring{$\mathbf{RQ_3}$}: What is the characterization of the origins of vulnerabilities and their pervasiveness within the {\tt npm} ecosystem's dependency network?}

Knowing the number of vulnerabilities within each parent package's dependency network enables us to analyse correlations between key variables, such as the number of dependencies and their impact on vulnerability exposure. The objective of $RQ_3$ is to examine these correlations and identify specific factors that developers can modify to minimize the risk of inheriting vulnerabilities through dependency chains.

\begin{figure}[htp!]
    \centering
    \includegraphics[width=0.8\linewidth]{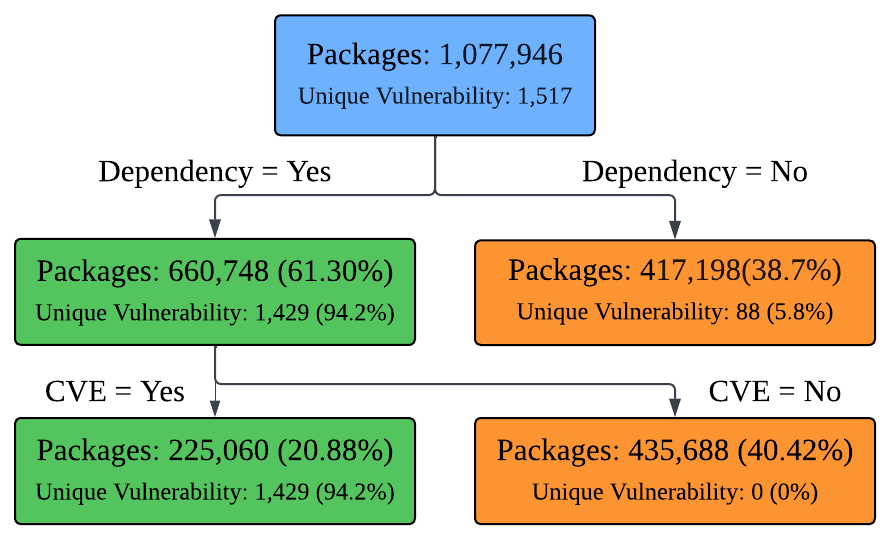}
    \caption{Distribution of software packages and unique vulnerabilities based on dependency status and presence of CVEs.}
    \label{fig:NPN_Dependency}
\end{figure}

Figure \ref{fig:NPN_Dependency} highlights the significant concentration of software vulnerabilities within packages that have dependencies. Out of the 1,077,946 packages analysed, 61.30\% possess one or more dependencies, accounting for a striking 94.2\% of all unique vulnerabilities. Furthermore, the presence of a vulnerability in dependency networks strongly correlates with an increased variety of vulnerabilities present, as 20.88\% of dependency packages with vulnerabilities present contain a disproportionate 94.2\% of all unique vulnerabilities. In contrast, packages without dependencies-—comprising 38.7\% of the total-—show a significantly lower vulnerability rate of just 5.8\%, with no unique vulnerabilities identified in standalone packages that lack vulnerabilities.

To analyse the dependency network of various {\tt npm} packages, we consider two directions of exploration based on the following two questions:
\begin{itemize} 
    \item [$RQ_{3.1}$]: To what extent do unique vulnerabilities (individual CVEs) spread across the software ecosystem as demonstrated by the number of affected packages?
    \item [$RQ_{3.2}$]: How do CVEs propagate across software ecosystems, and what role does dependency magnitude play in influencing their spread?
\end{itemize}

\subsubsection{Propagation Patterns of Software Vulnerabilities: Analysing the Spread and Influence of CVE Types Across Dependency Networks ($RQ_{3.1}$)}

\begin{figure}[ht!]
    \centering
     \includegraphics[width=0.85\linewidth]{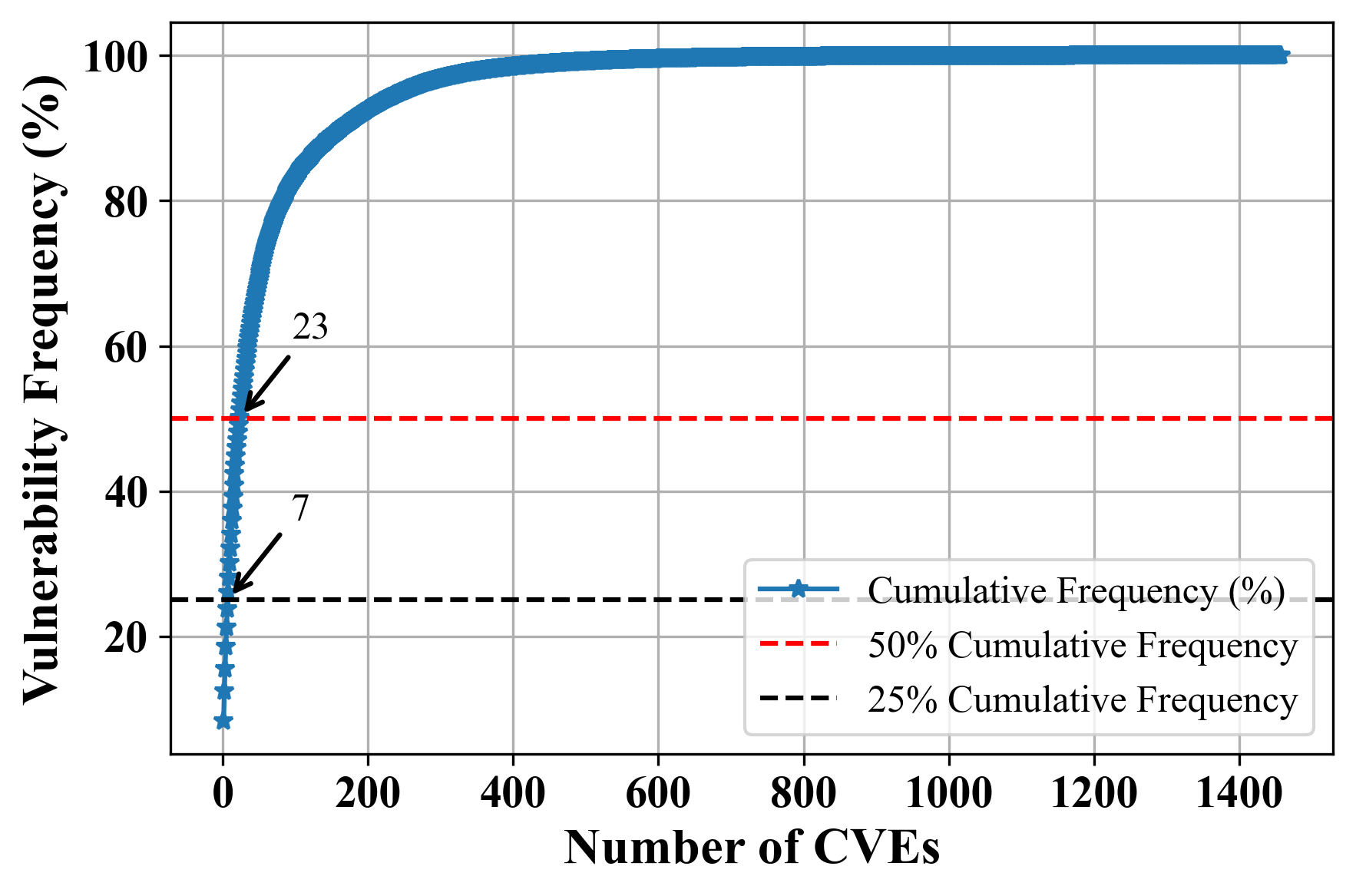}
     \vspace{-0.15cm}
    \caption{Distribution of CVE vulnerabilities.}
    \label{fig:dist_cves}
    \vspace{-0.3cm}
\end{figure}

Figure~\ref{fig:dist_cves} shows a steep initial rise in the curve, indicating that a small number of CVEs contribute to a large percentage of the 2,150,104 total vulnerability instances. Specifically, just five CVEs are responsible for 20\% (430,021 instances) of vulnerabilities, while approximately 23 unique CVEs account for 50\% (1,075,052 instances) of all observed vulnerabilities. This distribution highlights the potential impact of concentrating security efforts on these high-frequency CVEs, as addressing them could lead to a significant reduction in overall vulnerability risk.

Unique vulnerabilities can have a broad and cascading impact across the software ecosystem, often affecting multiple packages due to widespread code reuse, shared libraries, and transitive dependencies. Our studies have shown that a single CVE (\texttt{CVE-2023-44270}) is widely used in {\tt npm} libraries that can impact thousands of downstream packages. 
Details of these 23 CVEs and their distributors are shown in Figure \ref{fig:dist_cves_2} (Appendix \ref{CVEs_23}).

Table~\ref{tab:cves_impact} demonstrates how focusing remediation efforts on the most frequently occurring CVEs can substantially reduce ecosystem-wide vulnerability exposure. 
The Table lists the top-$N$ most frequent CVEs based on their occurrence in the dataset. Specifically, for the top-$N$ CVEs, it shows the number of times they appear, the total number of affected package versions, and the count of distinct packages impacted. Among the 225,060 distinct packages analysed, addressing the top-5 CVEs results in the elimination of 455,893 vulnerabilities, representing 21.24\% of the total reported vulnerabilities. This intervention affects 93,597 packages (41.58\%) and fully secures 19,136 distinct packages (8.50\%) by eradicating all known vulnerabilities within them. As the number of most-frequent CVEs increases, the benefits become increasingly pronounced. For instance, resolving the top-20 CVEs mitigates 1,009,584 vulnerabilities (47.04\%) and impacts 150,486 packages (66.86\%), while rendering 36,795 packages (16.35\%) free of vulnerabilities. 
By remediating the top-200 CVEs, over 92.68\% of the total vulnerabilities are resolved, and 73.55\% of the distinct packages are fully secured. 
\vspace{-2mm}\begin{table}[ht!]
    \centering
    \caption{Impact of Top-$N$ frequent CVEs on the number of vulnerabilities and affected packages.}
    \vspace{-0.15cm}
    \resizebox{0.98\linewidth}{!}{
    \begin{tabular}{c|c | c| c  }
    \toprule
      Top-$N$   &  Number of  & Number of Affected & Number of Distinct  \\
    CVEs & Vulnerabilities & Packages & Affected Packages \\ 
      \hline
      1 & 179,098 (8.34\%) & 24,882 (11.05\%) & 1,460 (0.65\%) \\ 
       5   & 455,893 (21.24\%)& 93,597 (41.58\%) & 19,136(8.50\%) \\
       10 & 688,797 (32.09\%) & 110,401 (49.05\%) & 20,444 (9.08\%)\\
       20 & 1,009,584 (47.04\%) & 150,486 (66.86\%) & 36,795 (16.35\%) \\
       50 & 1,487,092 (69.28\%) & 188,008 (83.53\%) & 86,787 (38.56\%)\\
        100 & 1,789,458 (83.37\%) & 202,254 (89.87\%) & 126,285 (56.11\%) \\ 
        200 & 1,989,100 (92.68\%) & 212,995 (94.64\%) & 165,536 (73.55\%) \\
    \bottomrule   
    \end{tabular}
    }
    \label{tab:cves_impact}
    \vspace{-0.3cm}
\end{table}
The results in Table \ref{tab:cves_impact} demonstrate that a limited number of high-frequency CVEs contribute to a large portion of vulnerabilities and affected packages, indicating their presence in widely reused dependency packages. This underscores the need for a detailed examination of packages that are frequently depended upon, as securing these can lead to significant vulnerability mitigation across the ecosystem due to their transitive impact.

\noindent\fbox{\parbox{0.99\linewidth}{
\textbf{{\textit Findings.}} We observe that 7 CVEs account for 25\% of the identified vulnerabilities, while 23 CVEs collectively contribute to 50\%. In particular, a single high-severity CVE (CVE-2023-44270) alone contributes 8.34\% of total vulnerabilities and affects 24,882 (11.05\%) packages. The top-10 CVEs impact 49.05\% of all affected packages. 
}} 


\subsubsection{Understanding CVE Propagation Across Software Ecosystems and Dependency Magnitudes ($RQ_{3.2})$}

\begin{figure}[h!]
    \centering
    \includegraphics[width=0.8\linewidth]{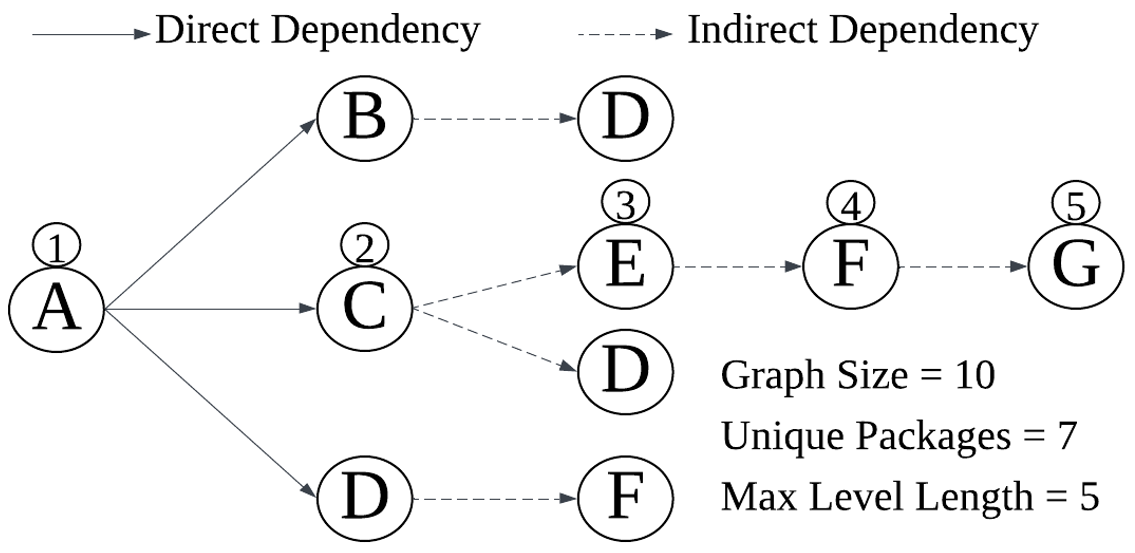}
    \caption{An example of a package dependency graph.  }
    \label{dep_graph}
    \vspace{-0.3cm}
\end{figure}

To understand CVE propagation in the {\tt npm} ecosystem, we need to analyse how the most frequent CVEs manifest throughout dependency networks. To achieve this, we quantify the total packages present throughout a given affected package's dependency network, and specifically each dependency 'level'.  Figure \ref{dep_graph} illustrates a sample dependency graph for package/node $A$. The graph consists of 10 nodes in total, indicating there exist 10 total packages. Among these, there exist 7 unique packages, and 5 dependency 'levels'. These metrics will be used for determining the propagation of CVEs throughout {\tt npm} dependency networks.

\begin{figure}[htp!]
    \centering
   \vspace{-2mm}\includegraphics[width=0.8\linewidth]{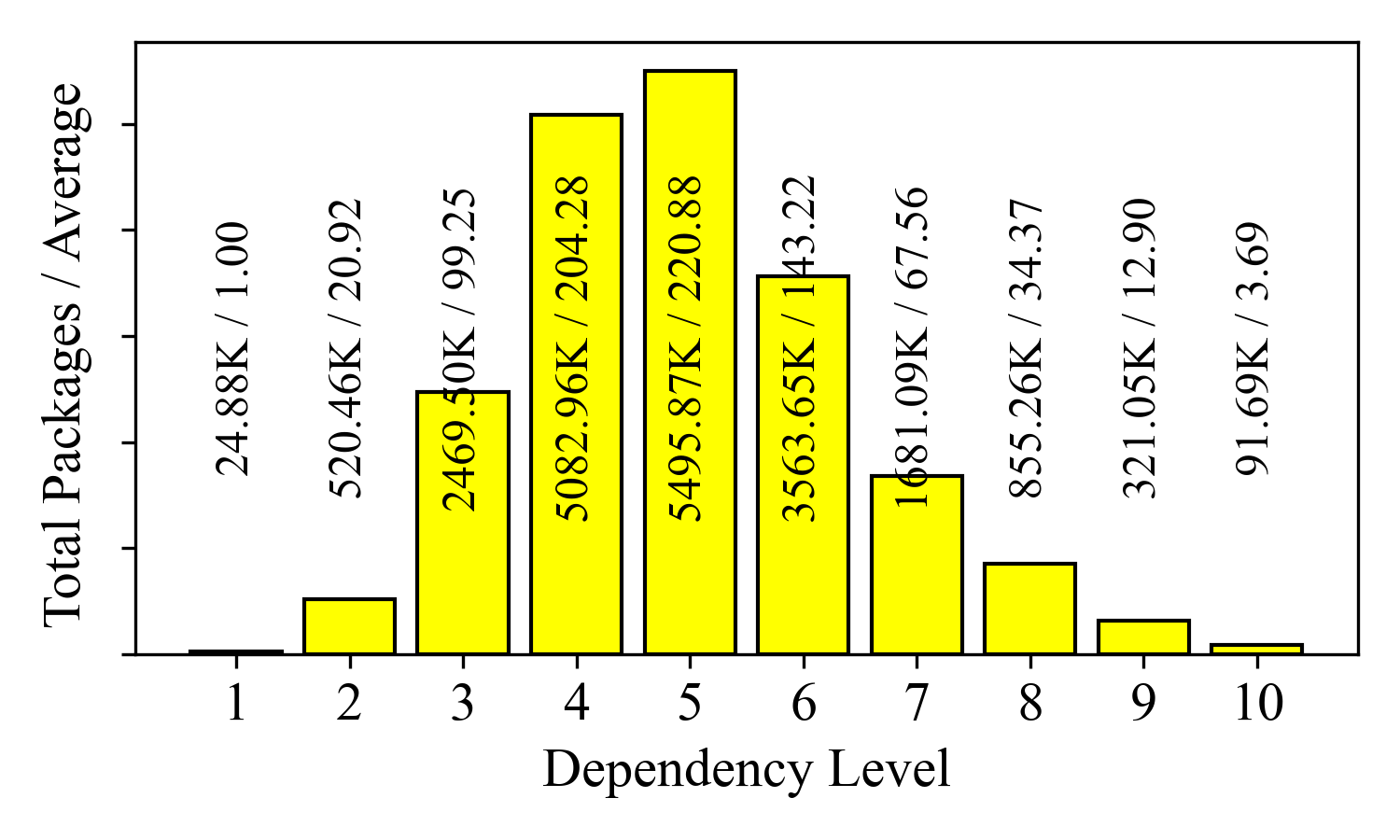} 
   \vspace{-0.15cm}
    \caption{Distribution of total and average affected packages by dependency level for \texttt{CVE-2023-44270}.}
    \label{fig:dependent_dist_for_top1}
    
\end{figure}

Firstly, we consider the most frequent vulnerability (\texttt{CVE-2023-44270}), which directly affects 24,882 packages.  Figure \ref{fig:dependent_dist_for_top1} shows the distribution of dependency packages by level for all parent packages affected by \texttt{CVE-2023-44270}. The $x$-axis represents the dependency level, and the $y$-axis shows the number of packages at each level. The distribution peaks at dependency levels ranging at a depth of 4-6 packages, indicating a highly interconnected ecosystem. This figure also depicts an identical trend for the average number of dependency packages per level per parent package. Similarly, this trend holds for the top-5, top-10, and top-20 CVEs (see Appendix \ref{appendix1}). 

From this, we may observe for CVE propagation that packages with higher dependency network depth are more susceptible to vulnerabilities, potentially leading to an increased spread. 
The concentration of packages with 4–6 dependencies suggests that a vulnerability in a common dependency could affect much of the ecosystem. Conversely, packages with fewer dependencies are more isolated from the risk of possessing vulnerabilities.
\begin{table*}[h!]
    \centering
    \def\arraystretch{1.25}
    \vspace{-0.25cm}
\caption{Spread of the most frequent CVEs across {\tt npm} packages and dependency networks.}
\vspace{-0.25cm}
    \resizebox{0.90\linewidth}{!}{
    \begin{tabular}{cccc  | c c c c}
    \toprule
      Top-$N$    &  Affected  &  Average No. & Dependency & \multicolumn{4}{|c}{Dependency Coverage (By No. Packages)}\\ \cline{5-8}
     CVEs & Packages &  Dependencies &  Length (Max.) & 5\% & 10\% & 25\% & 50\%\\ \hline 
       1  & 24,882 &  812.2 & 727 & 536 (2.2\%) & 1,195 (4.8\%) & 3,407 (13.7\%) & 7,425 (29.8\%)\\ 
       5 & 93,597 &  402.6 & 12.12 & 1,095 (1.2\%) & 2,432 (2.6\%) & 6,801 (7.3\%) &  16,110 (17.2\%)\\ 
       10 & 110,401 &  355.1 & 1.212 & 1,147 (1.0\%) & 2,544 (2.3\%) & 7,117 (6.4\%) & 17,091 (15.5\%)  \\ 
       20 & 150,486 & 298.1 & 1.212 & 1,339 (0.89\%) & 2,963 (2.0\%) & 8,338 (5.5\%) & 20,963 (13.9\%) \\ \bottomrule
    \end{tabular}
    }
    \label{tab:dependency_statistics}
    \vspace{-0.45cm}
\end{table*}
Table \ref{tab:dependency_statistics} summarizes how the packages affected by the top-$N$ CVEs ($N = 1$, $5$, $10$, $20$) propagate within the {\tt npm} ecosystem. For each group, the table reports the number of affected packages, the average number of dependencies per package, and the maximum dependency depth. Additionally, it quantifies the structural concentration of vulnerability spread by showing the percentage of affected packages that can be reached through the top 5\%, 10\%, 25\%, and 50\% of the most frequently used dependencies.
For example, a single CVE affected 24,882 packages, with an average of 812.2 dependencies per package and a maximum dependency depth of 727. The top 5\%, 10\%, 25\%, and 50\% of dependencies cover 2.2\%, 4.8\%, 13.7\%, and 29.8\% of the affected packages, respectively. As the number of CVEs increases, the number of affected packages also grows from 93,597 for the top-$5$ CVEs to 150,486 for the top-$20$ CVEs, while the average dependency size decreases from 402.6 to 298.1, respectively. However, the coverage percentages also decline, indicating a more dispersed vulnerability footprint. For instance, the top 50\% most used dependencies cover 17.2\% of affected packages for the top-$5$ CVEs, though drops to 13.9\% for the top-$20$.

\noindent\fbox{\parbox{0.99\linewidth}{
\textbf{{\textit Findings.}} {\tt npm} has a highly interconnected dependency network, with most packages at 4–6 levels deep and some CVEs impacting packages with over 800 dependencies on average. 
Notably, 13.9\% of packages, impacted by the top-$20$ CVEs, account for 50\% of the network coverage, indicating a vulnerability propagation ratio of 3.6 -- meaning each affected package influences, on average, 3.6 other packages through dependencies.}} 


\vspace{-2mm}\section{Discussion \& Recommendations}
\label{sec:discussion}

For package developers, maintaining up-to-date dependencies is crucial to minimising vulnerability exposure. Our dataset shows that outdated dependencies are widespread: out of 1,077,946 parent packages, 232,836 depend on at least one vulnerable package. Prior research similarly reports that 60\% of JavaScript projects rely on outdated dependencies~\cite{zerouali2018empirical}. To reduce this exposure, developers should adopt permissive version constraints rather than pinning a single version, ensuring that the latest minor or patch releases are installed without forcing disruptive major upgrades. Regular pruning of unused libraries and preventing development-only dependencies from leaking into production dependencies further reduces the attack surface.   
While it is important, it may also be risky to upgrade immediately to the latest version due to {\tt recency risks}:  newly released packages may not yet have undergone sufficient security auditing.  Missing vulnerabilities--whether undiscovered or intentionally obscured--are inevitable in rapidly evolving ecosystems. The most practical defence is therefore to reduce unnecessary external dependencies and maintain a clear separation between development and runtime dependencies.

\noindent{\bf Patch adoption behaviour.} Although our work primarily evaluates vulnerability prevalence and ecosystem impact, our dataset provides additional insights into remediation latency. The median time for a vulnerability to be fixed from its first affected release is {59 months} (4 years and 11 months). However, fixes are typically available \textit{before} publication: {83.1\%} of vulnerabilities have a fix released prior to their OSV/NVD publication date, with a median publication-to-fix offset of {--19 days}. Despite this, downstream adoption remains slow. Many dependent packages continue using outdated versions long after fixes are available, reinforcing that patch publication and patch adoption are distinct problems. 

This adoption gap aligns with ecosystem-wide evidence. For example, recent work shows that {96\%} of downloads containing known vulnerabilities occurred despite fixed versions being available~\cite{sonatype2024}. Similarly, Cox et al.\ found that \textbf{64.1\%} of dependencies had an update lag exceeding one year, with some delays extending up to eight years~\cite{cox2015measuring}. Combining these findings with our dependency-age profiling--which shows nearly a quarter of packages continuing to rely on vulnerable or outdated versions--indicates that \textit{slow patch adoption}, not slow patch publication, is a primary driver of prolonged exposure in the ecosystem.

\noindent{\bf Security culture and tooling.} Some packages without a history of disclosed vulnerabilities may be riskier than those with documented CVEs. A strong security culture, demonstrated through regular updates, transparent security reviews, and responsive patching behaviour, is often a better predictor of future security posture. Tools such as \texttt{npm audit}, which warn users during installation when a package is known to be vulnerable, remain crucial and should be actively integrated into CI/CD pipelines.

Package maintainers should prioritise discovering and fixing vulnerabilities before public disclosure, especially given the long-tail behaviour of vulnerabilities that remain unpatched for years. Security tooling (e.g., dependency analysis, update-suggestion systems, automated patching tools) should be improved to assist with timely identification and remediation.
Package managers should better surface deprecation notices and vulnerability warnings to reduce accidental continued use of outdated versions. Developers should monitor OSV/NVD feeds and adjust dependency constraints to ensure compatibility with patched releases. Consistent vulnerability scoring (e.g., adopting CVSS~v4~\cite{first2025}) should be encouraged to reduce ambiguity in prioritising fixes.

\noindent{\bf Network-scale insights.} Our analysis shows that the structure of the dependency graph greatly influences the impact of vulnerabilities. Only {20 CVEs} account for {half of all vulnerabilities} affecting interconnected packages (see Table~\ref{tab:dependency_statistics}). The average affected package influences {3.6 additional packages} via dependency chains, demonstrating the cascading nature of vulnerability propagation.

\noindent{\bf Limitations and future work.} Our analysis uses static dependency inheritance, which may overestimate exposure when vulnerable code is included but never executed. Conversely, dynamic analysis--though more precise--is computationally expensive at {\tt npm} scale. Additionally, approximately {400,000 packages} were removed or deprecated before data collection (see \S\ref{subsec:dsource}), potentially biasing results toward active libraries. Publication-date inconsistencies across OSV and NVD (averaging a {5-day discrepancy}, with some outliers exceeding {5 years}) introduced minor temporal noise (see Figure~\ref{fig:pubFromInitialCreation} in \S\ref{subsec:rq1}).
Future work should integrate static--dynamic analysis, quantify patch-adoption lag at individual dependency trees, and combine multiple vulnerability databases to improve date consistency.

\vspace{-3mm}\section{Threats to Validity}
\label{sec:threats}
Google OSV aggregates vulnerability reports from a variety of sources, such as {\tt snyk.io} and GitHub Advisories, providing a comprehensive but heterogeneous dataset. 
OSV.dev aggregates vulnerability reports from multiple open-source and ecosystem-specific sources — including GitHub Advisories — resulting in a comprehensive but heterogeneous dataset. This aggregation can introduce inconsistencies in the quality and detail of the reported vulnerabilities, as different sources may follow varying reporting standards and taxonomies. For example, GitHub advisories might focus on project-specific security issues, while {\tt snyk.io} could provide broader analyses that may not always align in terms of severity or impact. Consequently, some vulnerabilities could be underrepresented or overemphasized based on the source's reporting focus, leading to uneven insights into the security of dependencies across diverse ecosystems.

Additionally, as of November 2024, the industry is transitioning to CVSS version \texttt{4.0}. One identified issue with earlier versions, like CVSS \texttt{v3.x} is the potential for a small number of low-severity vulnerabilities to be overscored. This miscalculation often stems from the conversion of categorical attributes--such as the attack vector, complexity, and impact--into numerical severity ratings. 
Such overscoring can skew vulnerability-management priorities, especially in large dependency networks where low-risk issues inflate overall risk assessments.

CVSS \texttt{v4.0} is expected to address these issues by introducing an updated scoring schema. This new version refines the way categorical data is quantified, aiming to provide a more accurate reflection of real-world risk. For example, CVSS \texttt{v4.0} introduces greater granularity in the assessment of attack complexity and privileges required, which may result in a more precise rating of vulnerabilities. Furthermore, by improving the adaptability of the scoring mechanism to different contexts (e.g., cloud vs. on-premise environments), CVSS \texttt{v4.0} could reduce the likelihood of overscoring minor vulnerabilities, thus enabling more effective prioritization of remediation efforts within large-scale dependency networks.


\vspace{-2mm}\section{Related Work}
\label{sec:rwork}
Prior research in this field has provided several key insights, particularly regarding the existence of vulnerabilities within various software repositories and their respective dependency networks. Specifically for {\tt npm}, it has been observed that the number of vulnerabilities is increasing and that these vulnerabilities are being disclosed faster than other software repositories \citep{zerouali2022impact}. Moreover, it has been found that in both the {\tt npm} and RubyGems ecosystems, the time to disclose vulnerabilities has been increasing over time \citep{zerouali2022impact}. Combined with the insight that transitive dependencies increase nearly proportionally with the number of direct dependencies on {\tt npm} \citep{kikas2017structure}, and the strong correlation between a higher number of dependencies and a higher number of vulnerabilities \citep{gkortzis2021software}, this suggests a significant expansion of vulnerabilities within dependency networks over time. Decan et al. also found that many of these dependencies were outdated by several months \citep{decan2018evolution}. Additionally, an earlier study demonstrated that limiting the number of dependencies and relying exclusively on “trusted” packages can help mitigate the presence of vulnerabilities in parent packages \citep{bogart2016break}. Furthermore, remediation times for vulnerabilities in open-source projects have increased dramatically from approximately 25 days in 2017 to over 300 days in 2024 \citep{sonatype2024}.
Decan et al.~\cite{decan2018impact} present an empirical analysis of 399 security reports collected over six years across the npm dependency network, which includes over 600k JavaScript packages. It examines when and how vulnerabilities are discovered and fixed, assesses their severity, and reports key findings and offers recommendations for package maintainers and tool developers to strengthen vulnerability management. Liu et al.~\cite{liu2022demystifying} proposed a knowledge-graph approach (DVGraph) to resolve large-scale dependencies and track transitive vulnerability propagation, conducting an ecosystem-wide study and providing mitigation strategies. Zimmermann et al.~\cite{zimmermann2019small} analysed npm security risks, highlighting dependency-based single points of failure, maintainer influence, and the persistence of vulnerable, unmaintained packages, and recommending ecosystem-wide mitigations. 

While these studies provide valuable insights into the general landscape of vulnerabilities in {\tt npm} and other ecosystems, they do not offer a detailed, systematic characterization of how vulnerabilities propagate through dependency networks, nor do they investigate the relationship between vulnerability disclosure timelines and package outdatedness at scale. Our work builds on this foundation by introducing a quantitative analysis of how vulnerabilities spread across {\tt npm} dependencies, assessing whether outdated packages exacerbate security risks. 
Unlike prior studies that focus on vulnerability growth or broad dependency trends, we: \textit{(i)} analyse vulnerability propagation in the npm ecosystem by tracing how vulnerabilities spread through dependency networks; \textit{(ii)} examine time-to-disclosure to assess whether widely used packages remain unpatched for extended periods; and \textit{(iii)} evaluate the pervasiveness of vulnerabilities across npm dependency networks to characterize ecosystem-wide risk.


\vspace{-3.5mm}\section{Conclusion}
\label{sec:conclusion}

This study provides critical insights into the propagation of vulnerabilities within the {\tt npm} ecosystem. Approximately 13.9\% of packages affected by the top-20 CVEs account for 50\% of network coverage, indicating a vulnerability propagation ratio of 3.6. This suggests that each affected package, on average, influences 3.6 other packages through dependencies. With 21.60\% of packages impacted by vulnerabilities, our findings underscore significant security challenges. 
These findings highlight the need for a paradigm shift in the {\tt npm} ecosystem's security approach, emphasizing early vulnerability detection, automated safe updates, and faster disclosure-to-patch timelines. 

\section*{Ethical Considerations} This study relies exclusively on publicly available data sources ({\tt npm-follower}, deps.dev, OSV, and NVD). No new vulnerabilities were discovered or exploited, no human subjects were involved, and no non-public data was accessed. All vulnerability identifiers discussed are already publicly catalogued.

\bibliographystyle{ACM-Reference-Format}
\bibliography{references}

\appendix

\section{Appendix}

\subsection{Vulnerability Database Creation Algorithm}
\label{appendix:algo}

In this section, we present a detailed explanation of our Algorithm \ref{algo1} for the data linkage process outlined in Section~\ref{sec:vuldatalinkage}. 
The Unique Vulnerability Database Creation algorithm processes an OSV dataset to generate a structured vulnerability database enriched with severity scores from the NVD API. First, we initialise the empty database in line 1. The process begins with loading the OSV dataset in line 2, then vulnerability information is extracted with Extract\_Vulnerabilities in line 3. The algorithm identifies vulnerable packages by filtering those with at least one reported vulnerability, forming the set $\mathcal{P}_{vuln}$ in line 4. Then, dependency relationships are computed by iterating through each vulnerable package and mapping its dependencies in an inverse relationship structure, aiding in impact analysis in lines 5-7. To enhance efficiency, we employ parallel processing with a ProcessPoolExecutor that extracts vulnerability details for each package and uses ExtractVulnDetails and appends them to the database in lines 8-11. 

\begin{algorithm}[ht!]
\small
    \caption{Unique Vulnerability Database Creation ($\mathcal{D}_{OSV}$, $\mathcal{NVD}$)
    }
    \label{algo1}
    \KwData{ $\mathcal{D}_{OSV}$: OSV vulnerability data, $\mathcal{NVD}$: NVD Database.}
    
    \KwResult{ $\mathcal{D}_{CVE}$ vulnerability database with severity scores}

     $\mathcal{D}_{vulnerabilities} \leftarrow \emptyset$ 
     
    $\mathcal{D}_{OSV} \leftarrow$ LoadJSON(OSV\_dataset\_file) 
    
    $\mathcal{V} \leftarrow$ Extract\_Vulnerabilities($\mathcal{D}_{OSV}$) 

    $\mathcal{P}_{vuln} \leftarrow \{p \in \mathcal{D}_{OSV} \mid \exists v \in p.vulnerabilities\}$ 

    \For{$p \in \mathcal{P}_{vuln}$}
    {
    
        \For{$d \in p.dependencies$}
        {
        
            $\mathcal{R}_{inv} \leftarrow \mathcal{R}_{inv} \cup \{(d, p)\}$ 
            
        }
        
    }

   
    \textit{executor} $\leftarrow$ ProcessPoolExecutor() 
    
    \For{$p \in \mathcal{P}_{vuln}$ \textbf{in parallel with} \textit{executor}}
    {

        $v_{details} \leftarrow$ ExtractVulnDetails($p$) \tcp*{Extract relevant vulnerability details}
        
        $\mathcal{D}_{vulnerabilities} \leftarrow \mathcal{D}_{vulnerabilities} \cup v_{details}$ \tcp*{Append processed data}
        

    }

    
    \For{$batch \in$ Batches($\mathcal{D}_{vulnerabilities}, batch\_size$)}
    {

        $\mathcal{NVD}_{scores} \leftarrow$ FetchNVDscores($batch$) 
        
        \For{$cve \in batch$}
        {
            
            $s\_{severity} \leftarrow \mathcal{NVD}_{scores}[cve].baseSeverity$ 
            
            $s\_{base} \leftarrow \mathcal{NVD}_{scores}[cve].baseScore$ 
            
            $s\_{exploitability} \leftarrow \mathcal{NVD}_{scores}[cve].exploitabilityScore$ 
            
            $s\_{impact} \leftarrow \mathcal{NVD}_{scores}[cve].impactScore$ 
        }
        
        Wait(API\_rate\_limit) \tcp*{Comply with API rate limits}
        SaveCheckpoint($\mathcal{D}_{vulnerabilities}$) \tcp*{Save progress}
   }

    $\mathcal{D}_{CVE} \leftarrow \{cve \in \mathcal{D}_{vulnerabilities} \mid cve.status \neq$ ``Rejected''$\}$ 
    
    SaveJSON($\mathcal{D}_{CVE}$, output\_file) 

    \Return $\mathcal{D}_{CVE}$ \tcp*{Complete unique vulnerability database}
\end{algorithm}

After that, the algorithm fetches severity scores from the NVD API in batches, retrieving key metrics such as base severity, base score, exploitability score, and impact score in lines 12-18. It introduces a wait mechanism to comply with API rate limits while periodically saving progress in lines 19-20. Finally, the final vulnerability database is constructed by filtering out rejected CVEs before saving the cleaned dataset as a JSON file in lines 21-22 and returning the unique CVE database in line 23. The algorithm’s use of parallel execution, dependency tracking, and API-based enrichment ensures scalability and efficiency in unique vulnerability database construction.

\subsection{Distribution of Top 23 CVEs.}
\label{CVEs_23}
\begin{figure}[h!]
    \centering
    \includegraphics[width=0.95\linewidth]{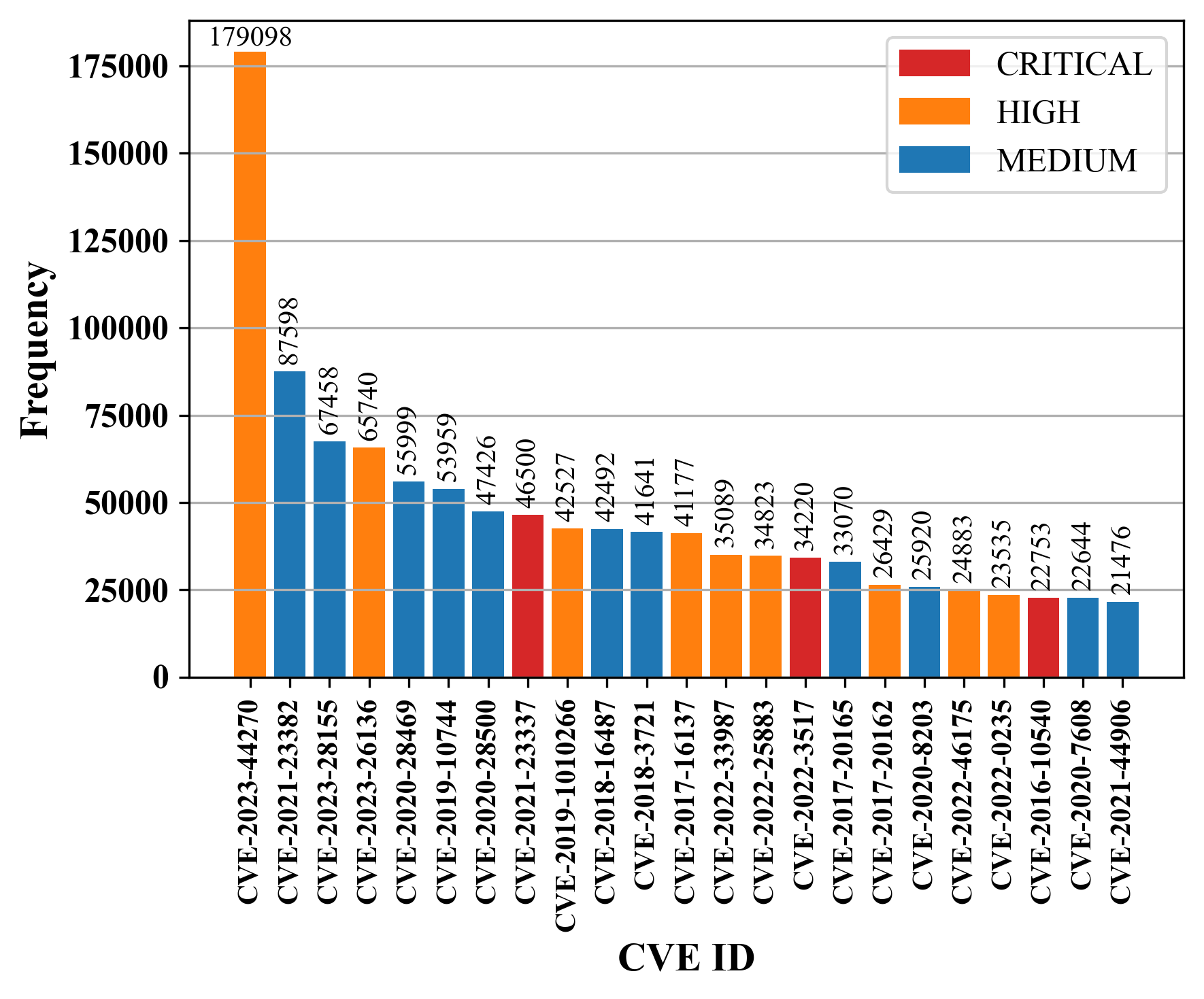}
    \caption{Distribution of top 23 CVE vulnerabilities frequency and severity.}
    \label{fig:dist_cves_2}
\end{figure}

Figure \ref{fig:dist_cves_2} illustrates the frequency distribution of the top 23 CVEs, categorized by severity. CVE-2023-44270 (HIGH) appears most frequently, with 179,098 occurrences—more than twice the count of the second most frequent CVE, CVE-2021-23382, which appears 87,598 times. Among these top 23 CVEs, 3 are classified as Critical, 9 as High, and 11 as Medium severity vulnerabilities.
The high occurrence of these CVEs, especially those marked as High and Critical, highlights their widespread impact across the software ecosystem. This underscores the urgent need for proactive mitigation strategies and timely vulnerability disclosure to minimize potential exploitation. 

\begin{figure*}[h!]
    \centering
    \subfloat[Top 5 CVE's affected package dependency graph.]{\includegraphics[width=0.45\linewidth]{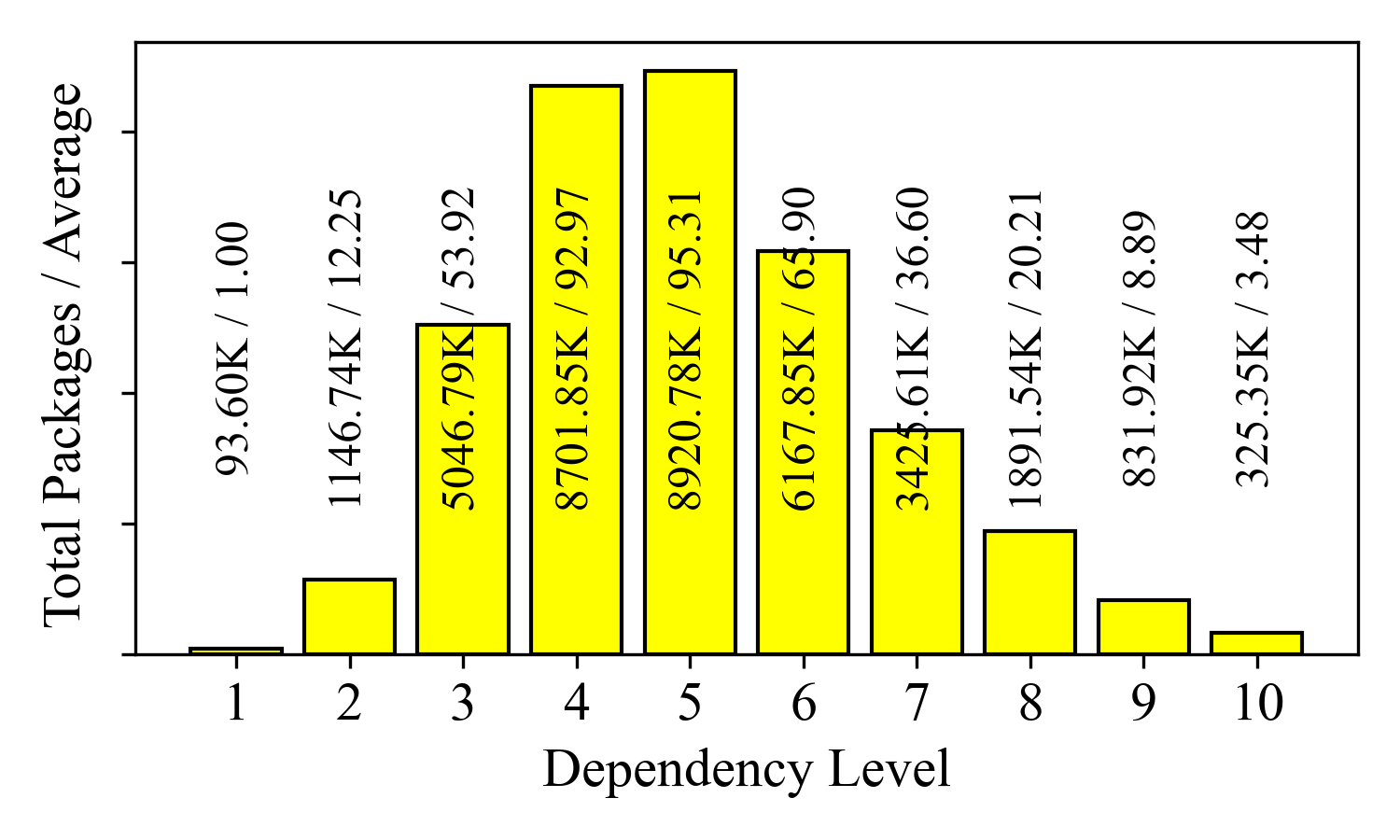}} 
    \subfloat[Top 10 CVE's affected package dependency graph.]{\includegraphics[width = 0.45\linewidth]{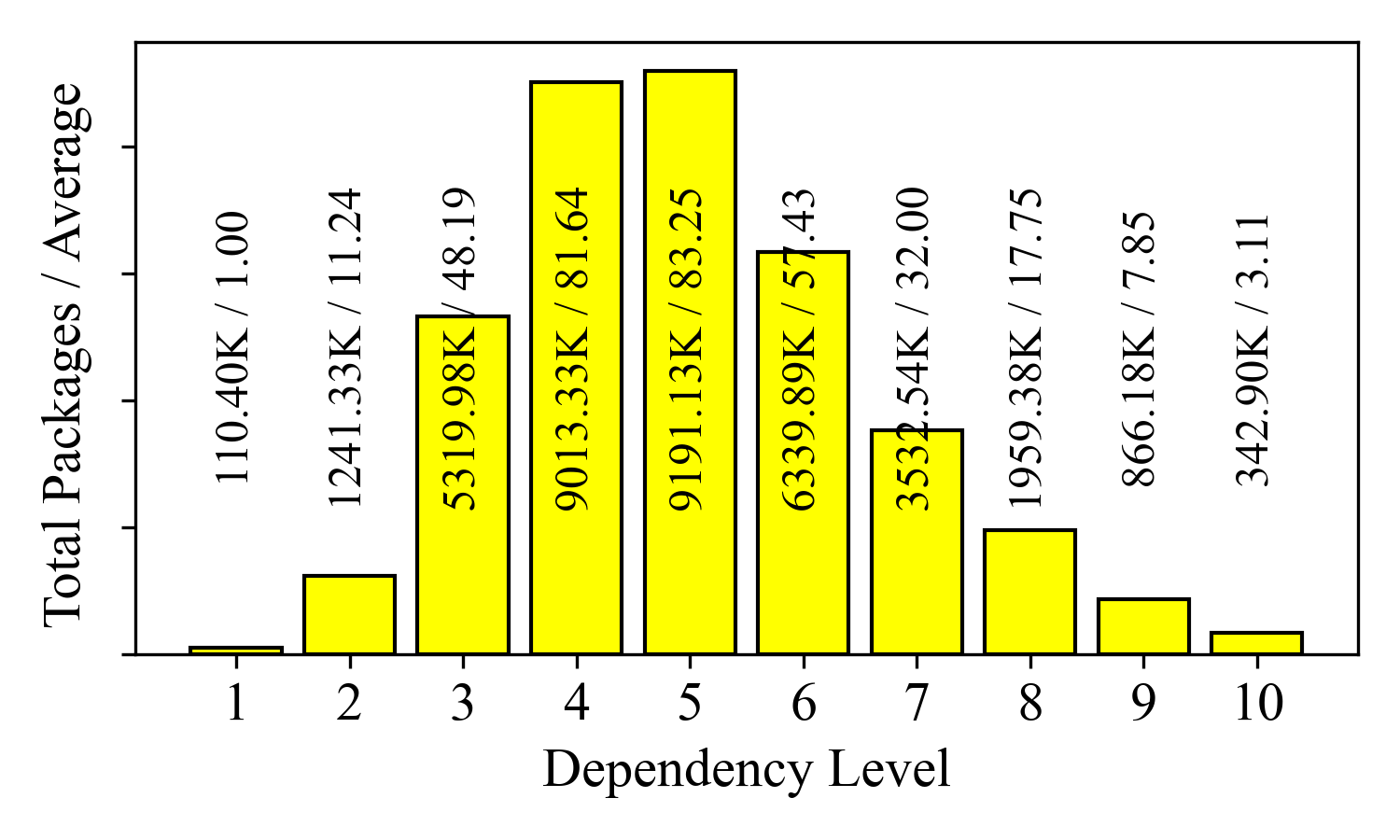}}\\
     \subfloat[Top 20 CVE's affected package dependency graph.]{\includegraphics[width = 0.45\linewidth]{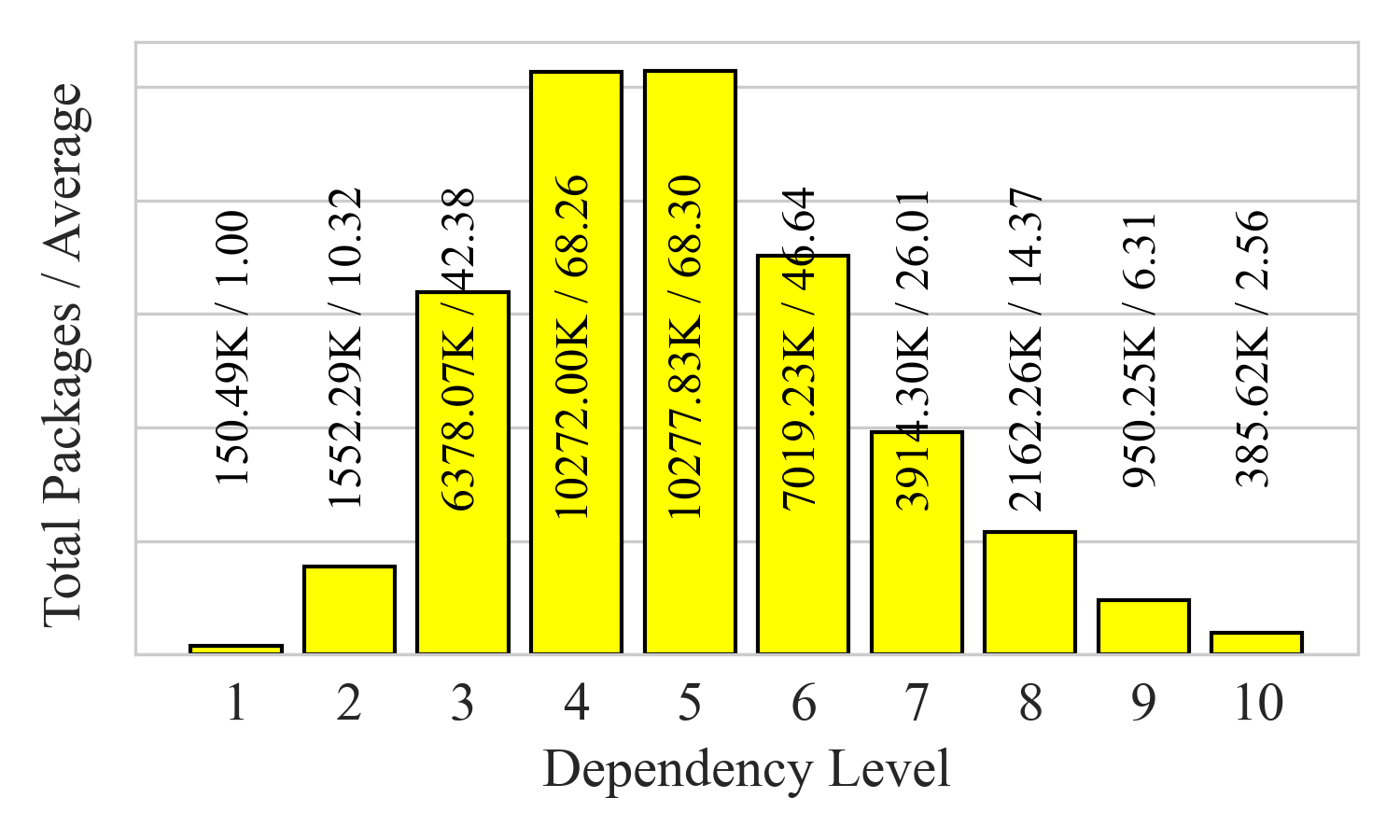}}
    \caption{Total number and average distribution of packages across dependency levels for top-5, top-10, and top-20 CVEs based on affected packages.}
    \label{fig:dependent_dist_for_top5}
\end{figure*}

\subsection{Dependency-level distribution of packages affected by the top 5, top 10, and top 20 CVEs} 
\label{appendix1}

Figure \ref{fig:dependent_dist_for_top5} presents the total count and average distribution of packages across dependency levels for the top 5, top 10, and top 20 CVEs.

\end{document}